\documentclass[preprint]{elsarticle}
\usepackage{amsmath}
\usepackage{amsfonts,amssymb}
\usepackage{bm}
\usepackage{graphicx}
\usepackage{array}
\usepackage{tabularx}
\usepackage{rotating}
\DeclareMathOperator*{\id}{id}
\DeclareMathOperator{\im}{im}
\DeclareMathOperator{\diag}{diag}

\newtheorem{thm}{Theorem}
\newtheorem{lem}{Lemma}
\newtheorem{cor}{Corollary}
\newdefinition{defi}{Definition}
\newdefinition{rmk}{Remark}
\newproof{pf}{Proof}
\begin{document}
 \title{Abe homotopy classification of topological excitations under the topological influence of vortices}
 \author[Phys.Univ.Tokyo,Keio.Univ]{Shingo Kobayashi}
 \author[Basic.Univ.Tokyo]{Michikazu Kobayashi}
 \author[Phys.Univ.Tokyo]{Yuki Kawaguchi}
 \author[Keio.Univ]{Muneto Nitta}
 \author[Phys.Univ.Tokyo,ERATO]{Masahito Ueda}
\address[Phys.Univ.Tokyo]{Department of Physics, University of Tokyo, 7-3-1 Hongo, Bunkyo-ku, Tokyo 113-0033, Japan }
\address[Keio.Univ]{Department of Physics, and Research and Education Center for Natural Sciences, Keio University 4-1-1 Hiyoshi, Yokohama, Kanagawa 223-8511}
\address[Basic.Univ.Tokyo]{Department of Basic Science, University of Tokyo, 3-8-1, Komaba, Meguro-ku, Tokyo, 153-8902, Japan}
\address[ERATO]{ERATO Macroscopic Quantum Control Project, JST, 7-3-1 Hongo, Bunkyo-ku, Tokyo 113-8656, Japan}

\begin{abstract}
Topological excitations are usually classified by the $n$th homotopy group $\pi_n$. However, for topological excitations that coexist with vortices, there are case in which an element of $\pi_n$ cannot properly describe the charge of a topological excitation due to the influence of the vortices. This is because an element of $\pi_n$ corresponding to the charge of a topological excitation may change when the topological excitation circumnavigates a vortex. This phenomenon is referred to as the action of $\pi_1$ on $\pi_n$. In this paper, we show that topological excitations coexisting with vortices are classified by the Abe homotopy group $\kappa_n$. The $n$th Abe homotopy group $\kappa_n$ is defined as a semi-direct product of $\pi_1$ and $\pi_n$. In this framework, the action of $\pi_1$ on $\pi_n$ is understood as originating from noncommutativity between $\pi_1$ and $\pi_n$. We show that a physical charge of a topological excitation can be described in terms of the conjugacy class of the Abe homotopy group. Moreover, the Abe homotopy group naturally describes vortex-pair creation and annihilation processes, which also influence topological excitations. We calculate the influence of vortices on topological excitations for the case in which the order parameter manifold is $S^n/K$, where $S^n$ is an $n$-dimensional sphere and $K$ is a discrete subgroup of $SO(n+1)$. We show that the influence of vortices on a topological excitation exists only if $n$ is even and $K$ includes a nontrivial element of $O(n)/SO(n)$.

\end{abstract}
\maketitle
\section{Introduction}
\label{intro}
Topological excitations exist in various subfields of physics, such as condensed matter physics, elementary particle physics, and cosmology. They have been  observed experimentally in superfluid helium systems, gaseous Bose-Einstein condensates (BECs), and liquid crystals. Topological excitations appear in ordered states, where the symmetry group $G$ of the  system reduces to its subgroup $H$ by spontaneous symmetry breaking. In an ordered state below the transition temperature, a system is characterized by the order parameter and is degenerate over the order parameter manifold $\mathcal{M} \simeq G/H$, $``\simeq"$ shows the relation of {\it homeomorphism}. The homotopy group is invariant under homeomorphism. In the low-energy effective theory, the order parameter varies in space within $\mathcal{M}$, so that the spatial variation of the order parameter costs only the kinetic energy. Topological excitation is defined as a nontrivial texture or singularity that is stable under an arbitrary continuous transformation in $\mathcal{M}$. 

Topological excitation is usually classified by homotopy theory \cite{Mermin;1979,Trebin;1982,Michel;1980,Mineev;1998,Volovik;2003}. The homotopy group not only determines the charges of topological excitations, but also stipulates the rules of coalescence and disintegration of topological excitations. For example, in the case of scalar BECs or superfluid helium 4, the symmetry group $G = U(1)$ reduces to $H = \{ 1\}$, and the order parameter manifold is $G/H \simeq U(1)$. Because $\pi_1 (U(1)) \cong \mathbb{Z}$, vortices created in these systems are characterized by integers, where $``\cong"$ shows the relation of {\it isomorphism}, and $\mathbb{Z}$ is an additive group of integers which implies that two vortices with winding number $1$ can continuously coalesce into a vortex with winding number $2$.  

The $n$th homotopy group $\pi_n$ can classify topological excitations with the dimension of homotopy $n = d -\nu -1$ $(n=d -\nu )$ for singular (nonsingular) topological excitations, where $d$ is the spatial dimension and $\nu$ is the dimension of topological excitation. Examples of singular topological excitations include a domain wall, a vortex, and a monopole, while those of nonsingular ones include a Sine-Gordon kink, a two-dimensional skyrmion, and a three-dimensional skyrmion. This classification is valid only when there is no other topological excitation with different dimensions of homotopy. In the case of $d =3$, however, when a vortex ($n=1$) and a monopole ($n=2$) coexist, $\pi_2$ is not always applicable for labeling the monopole.  Such a situation may occur in Kibble-Zurek phenomena \cite{Kibble;1976,Zurek;1985}. When we rapidly cool the system from a disordered to an ordered phase, topological excitations emerge spontaneously, because order parameters far away from each other are causally disconnected and can grow independently. Hence, it is possible for vortices, monopoles, and skyrmions to arise simultaneously and coexist. In such a case, topological excitations can influence each other and the very concept of the topological charge needs to be carefully reexamined. In this paper, we consider the situation in which topological excitations with the dimension of homotopy $1$ and $n$ ($\ge 2$) coexist. A nontrivial influence between a vortex and a monopole was first discussed by Volovik and Mineev \cite{Volovik;1977inf}, and Mermin \cite{Mermin;1978inf}. They have shown that a nontrivial influence of a vortex on a monopole exists for a nematic phase in a liquid crystal and a dipole-free state in a superfluid $^3$He-A phase. 
\begin{figure}[tbp]
\centering
 \includegraphics[width=8cm]{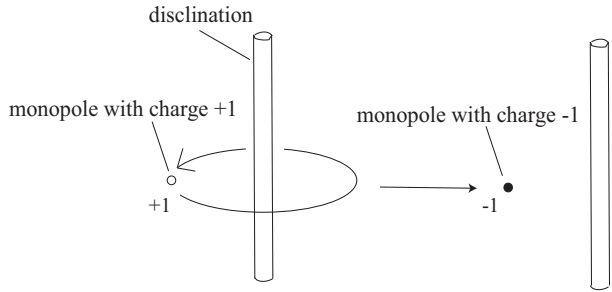}
 \caption{A monopole with charge $+1$ transforms to a monopole with charge $-1$ by making a complete circuit around a disclination \cite{Volovik;1977inf,Mermin;1978inf}.}\label{fig-0}
\end{figure}
For the case of a nematic phase in a liquid crystal, there is nontrivial influence of a disclination (a half quantum vortex) on a monopole. This influence is physically interpreted as a monopole with a charge $1$ changing its sign after rotating around the disclination (see Fig. \ref{fig-0}). Thus, if the disclination coexists with a monopole, the monopole charge defined by $\pi_2$ is no longer a  topological invariant. This effect is to be understood as the $\pi_1$ action on $\pi_2$. The influence of a vortex has been discussed in the context of classification of topological excitations in Refs. \cite{Mermin;1979,Trebin;1982,Michel;1980,Mineev;1998,Volovik;2003}.

In high-energy physics, Schwarz showed that symmetry breaking $SO(3) \to \mathbb{Z}_2 \ltimes SO(2)$ causes an influence of an Alice string on a monopole~\cite{Schwarz;1982}. Here, $``\ltimes"$ is a left semi-direct product, which means that $\mathbb{Z}_2$ nontrivially acts on $SO(2)$. Bucher, et al. studied the dynamics of the Cheshire charge that coexists with an Alice string~\cite{Bucher;1992}. 

In this paper, we use the Abe homotopy group~\cite{Abe;1940} to incorporate the influence of vortices on a monopole in the classification of topological excitations. We also show that the Abe homotopy group can classify all topological excitations under the influence of vortices. The $n$th Abe homotopy group is composed of $\pi_1$ and $\pi_n$. If elements of $\pi_1$ and $\pi_n$ do not commute, a nontrivial influence of vortices is shown to exist. The noncommutativity between $\pi_1$ and $\pi_n$ can naturally be incorporated into the structure of the Abe homotopy group. 
From the property of the Abe homotopy group, we show that the conjugacy class of the Abe homtopy group gives the physical charge of the topological excitation. The noncommuntativity between elements of $\pi_1 $ and $\pi_n$ implies that the Abe homotopy group is isomorphic to a nontrivial semi-direct product of $\pi_1$ and $\pi_n$.

We also calculate the influence of vortices for the case, in which $\mathcal{M} \simeq SO(n+1)/(SO(n) \rtimes K) \simeq S^n/K$, where $ S^n$ is an $n$-dimensional sphere embedded in $\mathbb{R}^{n+1} $ and $K$ is a discrete subgroup of $SO(n+1)$. Using Eilenberg's theory \cite{Eilenberg;1939}, we show that there is a nontrivial influence of vortices on topological excitations with homotopy dimension $n \ge 2$ if and only if $n$ is even and $K$ includes the nontrivial element of $O(n)/SO(n)$. 

This paper is organized as follows.
In Sec. \ref{sec:review}, we define notations used in this paper, review homotopy theory in Secs. \ref{sec:homotopy} -- \ref{sec:abe-homotopy}, and apply the Abe homotopy group to the classification of topological excitaions in Sec. \ref{sec:infl-Abe}. 
In Sec. \ref{sec:calc}, we calculate the influence of vortices by using Eilenberg's theory \cite{Eilenberg;1939}. 
In Sec. \ref{sec:app}, we apply the Abe homotopy group to liquid crystals, gaseous BECs, and superfluid helium systems. We confirm that an influence of vortices exists in the case of the uniaxial nematic phase in a liquid crystal, the polar phase in spin-1 BECs, the uniaxial nematic phase in spin-2 BECs, and the dipole-free A phase in a superfluid $^3$He. We find a nontrivial influence of vortices on an instanton classified by $\pi_4$ for the nematic phase in spin-2 BECs.
In Sec. \ref{sec:sum}, we summarize the main results of this paper.

\section{Homotopy theory and topological excitations}
\label{sec:review}
\subsection{Homotopy theory}
\label{sec:homotopy}
 Let us consider an ordered phase described by an order parameter $\phi$. A set of $\phi$ constitutes the order parameter manifold $\mathcal{M} = \{ g \phi | ^{\forall} g \in G/H\} \simeq G/H$. Here, $G$ is a group of operations that act on $\phi$ and do not change the free energy of the system, $H$ is an isotropy group of $G$ $(i. e.,  H = \{ g \in G | g \phi = \phi\})$, which is a subgroup of $G$ and whose elements leave $\phi$ invariant. In general, an excitation which is accompanied with a topological invariant is called a topological excitation; it is invariant under transformations of $G/H$.       

 Homotopy theory distinguishes whether an excitation is stable or not under continuous transformations. Two topological excitations are called homotopic if and only if there exists a continuous transformation between them. Homotopy equivalence classifies mappings from $S^n $ to $\mathcal{M}$, and the resulting equivalence classes are characterized by topological invariants. For example, in three-dimensional real space $\mathbb{R}^3$, a vortex is a line defect and a monopole is a point defect. We can judge whether they are stable or not by the maps $\phi|S^1$ and $\phi|S^2$, respectively. Here, $\phi|S^n \ \ (n \le 3)$ indicates that a domain of $\phi$ is restricted from $\mathbb{R}^3$ to $S^n$. A set of the mapping from $(S^n, s)$ to $(\mathcal{M}, \phi_0)$ is denoted by $(\mathcal{M}, \phi_0)^{(S^n, s)}$, which is called a functional space. Here, $\phi_0 \in \mathcal{M}$, which is defined by $ \phi_0 = \phi (s)$, is called a base point. A homotopy equivalent class of $f \in (\mathcal{M}, \phi_0)^{(S^1, s)}$ is denoted by $[f]$. A set of equivalent classes is written by $\pi_1 (\mathcal{M}, \phi_0)$, which features a group structure and is called a fundamental group. We can also construct a homotopy equivalent class of $(\mathcal{M}, \phi_0)^{(S^n,s)}$, which satisfies the group properties. A group of the equivalent classes of $(\mathcal{M}, \phi_0)^{(S^n, s)}$ is denoted by $\pi_n (\mathcal{M}, \phi_0)$ and is called the $n$th homotopy group. The stability of topological excitations is investigated by calculating $\pi_n (\mathcal{M}, \phi_0)$.  
\begin{figure}[tbp]
\centering
\includegraphics[width=8cm]{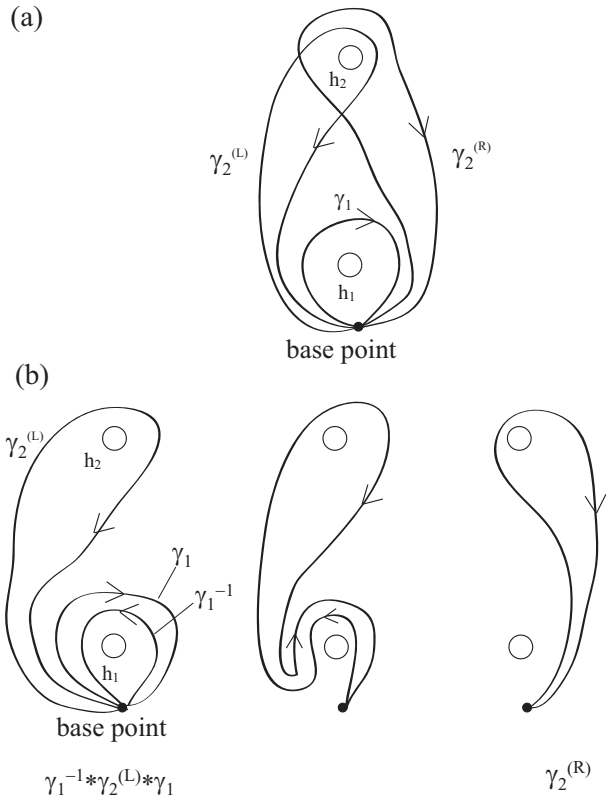}
\caption{ (a) Schematic illustration of closed loops in the order parameter manifold with two holes $h_1$ and  $h_2$, where $\gamma_1$ and $ \gamma_2$ are the loops enclosing $h_1$ and $ h_2$, respectively, in the clockwise direction. Here, the loop $\gamma_2$ that passes through the hole $h_1$ on the left-hand (right-hand) side is labeled $\gamma_2^{(L)}  (\gamma_2^{(R)})$. (b) Continuous deformations showing the equivalence of $\gamma_1^{-1} \ast \gamma^{(L)}_2 \ast \gamma_1 $ to $\gamma_2^{(R)}$.} \label{fig-1}
\end{figure}

If the homotopy group has an element other than the identity element, there exists a nontrivial topological excitation. Elements of a higher homotopy group $\pi_n (\mathcal{M}, \phi_0) \ \ (n \ge 2)$ always commute with each other, whereas elements of the fundamental group $\pi_1 (\mathcal{M}, \phi_0)$, in general, do not commute. If elements of the fundamental group do not commute each other, the corresponding two vortices create a rung structure upon collisions \cite{Mermin;1979,Michikazu;2009}. 

Consider two holes $h_1$ and $h_2$, each of them represents a vortex, in the order parameter space as shown in Fig.~\ref{fig-1}. The loops $\gamma_1$ and $\gamma_2^{(L,R)}$ enclose holes $h_1$ and $h_2$, respectively, in the clockwise direction. Here, $\gamma_2^{(L)}$ ($\gamma_2^{(R)}$) does not enclose the hole $h_1$ but goes through the left-hand (right-hand) side of $h_1$. We define a loop product
\begin{equation}
 f \ast g (t) = \begin{cases} f(2t) \; &0 \le t \le 1/2; \\ 
                                        g(2t -1) \; &1/2 \le t \le 1,  \end{cases} \label{S2-0}
\end{equation}
where $f$ and $g$ are maps satisfying $f:[0,1] \to (\mathcal{M},\phi_0), f(0) =f(1) = \phi_0$, and $g:[0,1] \to (\mathcal{M},\phi_0), g(0) =g(1) = \phi_0$. Then, one can easily verify that neither $\gamma_2^{(L)}$ nor $\gamma_2^{(R)}$ commutes with $\gamma_1$, i.e., $\gamma_2 ^{(L,R)}\ast \gamma_1 \neq \gamma_1 \ast \gamma_2^{(L,R)}$, but that they satisfy the relation $\gamma_1^{-1} \ast \gamma_2^{(L)} \ast \gamma_1 = \gamma_2^{(R)}$. Hence, $\gamma_2^{(R)}$ is not homotopically equivalent to $\gamma_2^{(L)}$. However, since both $\gamma_2^{(R)}$ and $\gamma_2^{(L)}$ enclose the same hole $h_2$, they should represent the same vortex. Thus, we impose the following equivalence relation on $\pi_1 (\mathcal{M}, \phi_0)$:
\begin{equation}
 \gamma_1^{-1} \ast \gamma_2^{(i)} \ast \gamma_1 \sim \gamma_2^{(i)  }, \ \ i=R,L. \label{S2-2}
\end{equation}
This equivalence relation is interpreted as a physical quantity being invariant under inner automorphisms on $\pi_1 (\mathcal{M}, \phi_0)$. Therefore, the conjugacy class of $\pi_1 (\mathcal{M}, \phi_0)$ classifies the type of vortices. If the fundamental group is commutative, each conjugacy class is composed of only one element.

\subsection{Homotopy group with base space $\mu$}
\label{sec:homotopy-exp}
We discuss the classification of topological excitations in a system that involves vortices. The order parameter changes continuously except at vortex cores. Now, we assume that topological excitations with the dimension of homotopy $n$ coexist with a vortex that does not directly interact with them. The topological excitations, however, are topologically affected by the vortex. In such a situation, we can no longer regard the element of $ \pi_n$ as a charge of the topological excitation. To define the topological invariant in the presence of vortices, we generalize the definition of the homotopy group using base space $\mu$.

\begin{figure}[tbp]
\centering
\includegraphics[width=10cm]{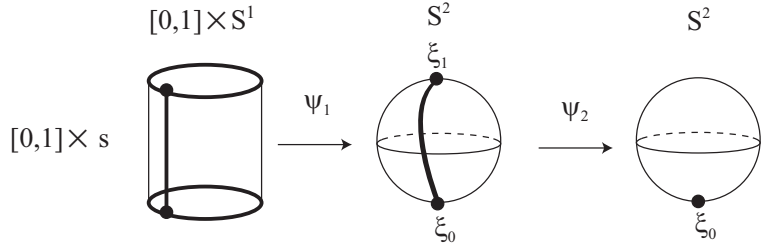}
\caption{
Mapping from $[0,1] \times S^{n-1}$ to $S^n$ for the case of $n=2$, where the domain $[0,1] \times S^1$ is a tube with a unit length. First, $\psi_1$ shrinks the top circle $ 1 \times S^1$ and the bottom circle $ 0 \times S^1$ to points $\xi_1$ and $\xi_0$, respectively, which involves the map from the path $[0,1] \times s $ to the path $\overline{\xi_0 \xi_1}$. Next, $\psi_2$ maps $\overline{\xi_0 \xi_1}$to $\xi_0$ and the complementary space $S^2 - \overline{\xi_0 \xi_1}$ to $S^n - \xi_0$. The bold curves and dots show the space which is mapped to $\phi_0 \in \mathcal{M}$ by $f^{\text{b}}$.
}\label{fig-2}
\end{figure}

First, we define the product of a path and a map of $(\mathcal{M}, \phi_0)^{(S^n,s)}$. Let us expand the domain $S^n$ to $[0,1] \times S^{n-1}$ and define a map $f^{\text{b}}$ as 
\begin{equation}
 f^{\text{b}} :([0,1] \times S^{n-1},  0 \times S^{n-1}  \cup  1 \times S^{n-1}  \cup [0,1] \times  s  ) \to (\mathcal{M}, \phi_0), \label{A2-1-1}
\end{equation}
where $0$ and $1$ are boundaries in $[0,1]$, $s$ is a point in $S^{n-1}$, and the subspace $0 \times S^{n-1}  \cup  1 \times S^{n-1}  \cup [0,1] \times s $ is mapped to $\phi_0$ by $f^{\text{b}}$.
That $f^{\text{b}}$ is isomorphic to $f \in (\mathcal{M}, \phi_0)^{(S^n,s)}$ can be shown as follows. We define a map $\psi_1$ as a map from $ 0 \times S^{n-1}  \cup  1 \times S^{n-1} $ to $\xi_0 \cup \xi_1$, where $\xi_0$ is an arbitrary point in $ 0 \times S^{n-1}$ and $\xi_1$ is an arbitrary point in $ 1  \times S^{n-1}$. Here, $\psi_1$ involves a map from a path $[0,1] \times  s$ to a path $ \overline{\xi_0 \xi_1}$. Here, $ \overline{\xi_0 \xi_1}$ is along a meridian of $S^n$ when we regard $\xi_0$ and $\xi_1$ as the south and north poles, respectively. Under the map $\psi_1$, $[0,1] \times S^{n-1}$ is mapped to $S^n$. Next, we define a map $\psi_2$ which maps the path $\overline{\xi_0 \xi_1}$ and its complementary space $S^n - \overline{\xi_0 \xi_1}$ to $\xi_0$ and $S^n - \xi_0$, respectively. Figure \ref{fig-2} illustrate the maps $\psi_1$ and $\psi_2$ for the case of $n=2$.
Then, $f^{\text{b}}$ is related to $f$ as
\begin{equation}
 f^{\text{b}} = f \circ \psi_2 \circ \psi_1, \label{A2-1-2}
\end{equation} 
where $`` \circ "$ denotes the composition of maps: $f \circ g (x) := f(g(x))$. Here, $f^{\text{b}}$ and $f$ are related in a one to one correspondence under $\psi_2 \circ \psi_1$ and belong to the same homotopy equivalence class of $\pi_n (\mathcal{M}, \phi_0)$. 
Let $\eta: [0,1] \to \mathcal{M}$ be a path on $\mathcal{M}$ such that $\eta (0) = \phi_0$ and $\eta (1) = \phi_1$, where $\phi_0$ and $\phi_1$ are arbitrary points on $\mathcal{M}$. The product of the path and the map $f^{\text{b}}$ is defined as
\begin{align}
  f^{\text{b}} \ast \eta (t,x) := \begin{cases} f^{\text{b}} (2t, x) \, &\text{for } \, 0 \le t \le \frac{1}{2}, \,  ^{\forall} x \in S^{n-1}; \\ \eta (2 t -1) \, &\text{for } \, \frac{1}{2} \le t \le 1, \,  ^{\forall} x \in S^{n-1}. \end{cases} \label{prod-1}
\end{align}
Similarly, we can obtain the product in a reverse order $\eta \ast f^{\text{b}}$.  
The product of the map $f^{\text{b}}$ and a loop $l$, which satisfies $l (0) = l(1) = \phi_0$, is defined in a similar manner as
\begin{equation}
 f^{\text{b}} \ast l (t,x) := \begin{cases} f^{\text{b}} (2t, x) \, &\text{for } \, 0 \le t \le \frac{1}{2}, \, ^{\forall} x \in S^{n-1} ; \\ l (2 t -1) \, &\text{for } \, \frac{1}{2} \le t \le 1, \, ^{\forall} x \in S^{n-1}. \end{cases} \label{prod-2}
\end{equation}
We define the product of the elements $[f^{\text{b}}] \in \pi_n (\mathcal{M}, \phi_0)$ and $[l] \in \pi_1 (\mathcal{M}, \phi_0)$ as $[f^{\text{b}} ] \ast [l] = [f^{\text{b}} \ast l]$. In general, a homotopy equivalence class of the map $[f^{\text{b}} \ast l]$ belongs to an element of $\pi_n (\mathcal{M}, \phi_0)$ because $f^{\text{b}} \ast l$ is the same type of map as that defined in Eq. (\ref{A2-1-1}) which is isomorphic to an element of $(\mathcal{M}, \phi_0)^{(S^n,s)}$. We also define the product of elements of $\pi_n (\mathcal{M}, \phi_0)$. Let $f^{\text{b}}, g^{\text{b}}$ to be elements of $(\mathcal{M}, \phi)^{([0,1] \times S^{n-1},  0 \times S^{n-1}  \cup  1 \times S^{n-1}  \cup [0,1] \times  s  )}$. The product between $f^{\text{b}}$ and $g^{\text{b}}$ is defined as
\begin{equation}
 f^{\text{b}} \ast g^{\text{b}} (t,x):=  \begin{cases} f^{\text{b}} (2t, x) \, &\text{for } \, 0 \le t \le \frac{1}{2}, \, ^{\forall} x \in S^{n-1} ; \\ g^{\text{b}} (2 t -1,x) \, &\text{for } \, \frac{1}{2} \le t \le 1, \, ^{\forall} x \in S^{n-1}. \end{cases} \label{prod-3}
\end{equation} 
The product in $\pi_n (\mathcal{M}, \phi_0)$ is defined by $[f^{\text{b}} ] \ast [g^{\text{b}}] : = [f^{\text{b}} \ast g^{\text{b}}]$.
In the following discussion, we denote $[f] = \alpha \; (f \in (\mathcal{M}, \phi_0)^{(S^n,s)})$ as the element of $\pi_n (\mathcal{M}, \phi_0)$, whereas the product of $\pi_n (\mathcal{M}, \phi_0)$ is defined by Eqs.(\ref{prod-1}), (\ref{prod-2}), and (\ref{prod-3}) because of Eq. (\ref{A2-1-2}).  
 
\begin{figure}[tbp]
\centering
\includegraphics[width=8cm]{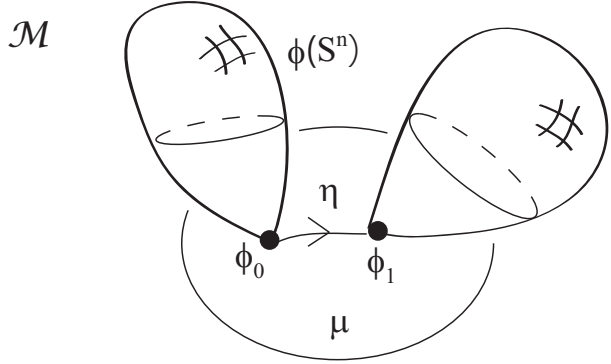}
\caption{
Schematic illustration showing the relationship between $\pi_n (\mathcal{M}, \phi_0)$ and $\pi_n (\mathcal{M}, \phi_1)$, where $\phi_0$ and $\phi_1$ are points in $\mu$ which is a subspace of $\mathcal{M}$.  $\pi_n (\mathcal{M}, \phi_0)$ is equivalent to $\pi_n (\mathcal{M}, \phi_1)$ if there is a path $\eta$ in $\mu$ that connects $\phi_0$ to $\phi_1$.
}\label{fig-3}
\end{figure}

\begin{defi}[The $n$th homotopy group with base space $\mu$]
Let $\mu$ be a connected subspace on $\mathcal{M}$, $ \phi_0$ be a point on $\mu$, and $f$ be a mapping from $(S^n,s)$ to $(\mathcal{M}, \phi_0)$. Then, we define a set of equivalent classes which have base points on $\mu$ as $\zeta_n (\mathcal{M},\mu) = \{ [f] \in \pi_n (\mathcal{M}, \phi) |   \ \ ^{\forall}\phi \in \mu\}$. For any $[f] \in \pi_n (\mathcal{M}, \phi_0) $ and $[g] \in \pi_n (\mathcal{M}, \phi_1)$, where $\phi_0 , \phi_1 \in \mu$, $[f]$ is equivalent to $[g]$ if there is a path $\eta: [0,1] \to \mu$ such that $\eta (0) = \phi_0,\eta(1)=\phi_1$, and
\begin{equation}
[g]  = [\eta]^{-1} \ast [f] \ast [\eta ], \label{S3-1}
\end{equation}
where $\ast$ is defined through Eq. (\ref{prod-1}). The equivalent classes of $\zeta_n (\mathcal{M},\mu)$ under Eq. (\ref{S3-1}) form a group which we write as $\pi_n (\mathcal{M} ,\mu)$. Figure \ref{fig-3} schematically shows the equivalence relation. We call $\pi_n(\mathcal{M},\mu)$ a homotopy group with a base space $\mu$. We write an element of $\pi_n (\mathcal{M} ,\mu)$ by $\{ f \}$ instead of $[f]$. If $\mu = \phi_0$, $\pi_n (\mathcal{M} ,\mu)$ reduces to $\pi_n (\mathcal{M} ,\phi_0)$ and $\{ f\} $ reduces to $[f]$.
\end{defi}

We define the product and the inverse in $\pi_n (\mathcal{M}, \mu)$. Let $\{ f\}$ and $\{ g\} $ be elements of $\pi_n (\mathcal{M}, \mu)$, and a map with a base point $\phi_0$ be $f_{\phi_0}$. The group structure of $\pi_n (\mathcal{M}, \mu)$ is induced by $\pi_n (\mathcal{M}, \phi_0)$. Namely, we define the product of $\pi_n (\mathcal{M}, \mu)$ as
\begin{equation}
 \{ f_{\phi_0} \} \ast \{ g_{\phi_0} \} =  \{[f_{\phi_0}] \ast [g_{\phi_0}] \}. \label{S3-1-2}
\end{equation}
Here, we can choose an arbitrary point on a base space $\mu $ as the base point because the element of $\pi_n (\mathcal{M}, \mu)$ satisfies 
\begin{equation}
   \{ f_{\phi_0} \} \ast \{ g_{\phi_0} \} = \{ f_{\phi_1} \} \ast \{ g_{\phi_1}\} \quad \text{for } \ \ ^{\forall}\phi_0 ,\phi_1 \in \mu ,  \label{S3-2}
\end{equation}
where $[ f_{\phi_1} ] = [\eta]^{-1} \ast [f_{\phi_0}] \ast [\eta]$ and $[g_{ \phi_1 } ] = [\eta]^{-1} \ast [ g_{\phi_0} ] \ast [\eta] $ with $\eta$ being a path on $\mu$ satisfying $\eta (0) = \phi_0$ and $\eta (1) = \phi_1$. From Eq. (\ref{S3-1-2}), we obtain 
\begin{equation}
 \{ f_{\phi_0}\} \ast \{ f_{\phi_0} ^{-1}\} =\{ 1_{\phi_0}\} \quad \text{for } \ \ ^{\forall} \phi_0 \in \mu,  \label{S3-3}
\end{equation}
where $1_{\phi_0}$ is a constant map defined by $1_{\phi_0} = \phi_0 \in \mu$ for $^{\forall} x \in S^n$.
Thus, the inverse of $\{f_{\phi_0}\}$ is given by $\{ f_{\phi_0} \}^{-1} = \{ f_{\phi_0}^{-1}\}$.
\begin{figure}[tbp]
\centering
 \includegraphics[width=8cm]{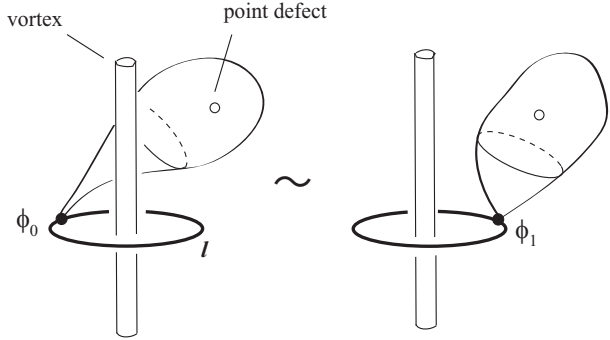}
 \caption{
 Equivalence relation for elements of $\pi_2 (\mathcal{M}, l)$, where $\sim $ denotes the equivalence and $l$ represents a loop encircling a vortex. The left element involves a closed surface that encloses a point defect with respect to a base point $\phi_0$ ($\phi_0$). These two elements are equivalent in $\pi_2 (\mathcal{M}, l)$ because $\phi_0$ is connected with $\phi_1$ along a path in $l$, and both closed surfaces enclose the same point defect. We define the charge of the point defect for the equivalent class of the closed surfaces. The charge thus defined constitutes a topological invariant. 
} \label{fig-4}
\end{figure}

The homotopy group with a base space $\mu $ includes the influence of vortices. Let $[l] = \gamma$ be an element of $\pi_1 (\mathcal{M}, \phi_0)$, which represents a vortex. [Precisely speaking, vortices are classified by the conjugacy classes of $\pi_1 (\mathcal{M}, \phi_0) $, as discussed in Sec. \ref{sec:homotopy}. ] 
When we choose a loop $l$ enclosing a vortex as a base space (see Fig. \ref{fig-4}), $\pi_n (\mathcal{M}, l)$ classifies topological excitations in the presence of a vortex $\gamma$, since 
topological excitations are independent of the choice of a base point on $l$.
By the definition of $\pi_n (\mathcal{M}, l)$, $[f_{l(0)}]$ and $[f_{l(t)}]$ are equivalent since there is a path $\eta_t :[0,1] \to l$ such that $\eta_t (s) = l(ts) \, $ for any $t,s \in [0,1]$ and $[f_{l(t)}] = \eta_t^{-1} \ast [f_{l(0)}] \ast \eta_t$. When $t < 1$, the equivalence relation $[f_{l(0)}] \sim [f_{l(t)}]$ gives a one-to-one correspondence between $\pi_n (\mathcal{M},l(0))$ and $\pi_n(\mathcal{M}, l(t))$. On the other hand, when $t=1$, since $l(0) = l(1) = \phi_0$, we obtain the equivalence relation
\begin{equation}
 [f_{\phi_0}] \sim \gamma ([f_{\phi_0}]), \label{S3-5}
\end{equation}
where
\begin{align}
 \gamma([f_{\phi_0}]) := \gamma^{-1} \ast [f_{\phi_0}] \ast \gamma. \label{S3-4}
\end{align}
Hence, $\pi_n(\mathcal{M}, l)$ is given by a set of the equivalent classes of $\pi_n (\mathcal{M}, \phi_0)$ under the equivalence relation (\ref{S3-5}) and it does not depend on the choice of the reference element $l$:
\begin{equation}
 \pi_n(\mathcal{M}, \gamma) := \pi_n (\mathcal{M}, l) \cong \pi_n ( \mathcal{M}, \phi_0 )/\sim. \label{S3-6}
\end{equation}
When the topological excitation returns to the original state after making a complete circuit of the loop, we have $[f_{\phi_0}] = \gamma^{-1} \ast [f_{\phi_0}] \ast \gamma $, and (\ref{S3-5}) is always satisfied. In this case, the homotopy group does not depend on the base point and thus we can represent the homotopy group without referring to the base point:
\begin{equation}
 \pi_n (\mathcal{M},\gamma) \cong \pi_n (\mathcal{M},\phi_0) \cong \pi_n (\mathcal{M}). \label{S3-7}
\end{equation}
If the action of $\pi_1(\mathcal{M}, \phi_0)$ is trivial, i.e., $\gamma ([f_{\phi_0}]) = [f_{\phi_0}]$ for any $\gamma \in \pi_1 (\mathcal{M}, \phi_0)$, the order parameter manifold $\mathcal{M}$ is called an $n$-simple \cite{Mermin;1979,Eilenberg;1939}. If the order parameter manifold is not an $n$-simple, the charge at the initial point $\alpha \in \pi_n (\mathcal{M},\phi_0)$ is, in general, different from that at the final point $\gamma(\alpha)$. Then, the charge should be defined as an element of $\pi_n (\mathcal{M}, \gamma)$.

As a result, a charge of a topological excitation that coexists with a vortex is characterized by an equivalence class of $\pi_n (\mathcal{M}, \phi_0)$ under the equivalent relation $[f_{\phi_0}] \sim \gamma([f_{\phi_0}])$, where $[f_{\phi_0}] \in \pi_n (\mathcal{M},\phi_0)$. The topological charge defined in this way is invariant even in the presence of vortices.

\subsection{Abe homotopy theory}
\label{sec:abe-homotopy}

In this section, we define the Abe homotopy group and explain its group structure. The description of topological excitations in terms of the Abe homotopy group \cite{Abe;1940} is given in the next subsection.  
As explained in Sec. \ref{sec:homotopy-exp}, a set of the homotopy equivalence class of maps from $([0,1] \times S^{n-1}, 0 \times S^{n-1}  \cup  1 \times S^{n-1}  \cup [0,1] \times  s )$ to $(\mathcal{M},\phi_0)$ is isomorphic to the $n$th homotopy group $\pi_n (\mathcal{M},\phi_0)$. 
The Abe homotopy group is defined as a set of homotopy equivalence class of maps from $([0,1] \times S^{n-1},   0 \times S^{n-1} \cup  1 \times S^{n-1} )$ to $(\mathcal{M} , \phi_0)$.  
\begin{defi}[The $n$th Abe homotopy group \cite{Abe;1940,Fox;1948}]
Let $\mathcal{M}$ be an arbitrary topological space and $f^{\text{a}}_{\phi_0}$ is a map such that 
\begin{equation}
f^{\rm a}_{\phi_0} : ([0,1] \times S^{n-1} ,  0 \times S^{n-1}  \cup 1 \times S^{n-1} ) \to ( \mathcal{M} ,\phi_0 ). \label{A2-1}
\end{equation}
 The subspace $ 0 \times S^{n-1} $ and $1 \times S^{n-1}$ are the top and bottom of the $n$th-dimensional cylinder, respectively, and they are mapped onto the base point $\phi_0$ by $f_{\phi_0}^{\rm a}$. A set of maps defined in (\ref{A2-1}) is denoted by $K_n (\mathcal{M}, \phi_0)$. The homotopy equivalence classes of $K_n (\mathcal{M}, \phi_0)$ constitute a group, which is denoted by $\kappa_n (\mathcal{M}, \phi_0)$. 
\end{defi}

\begin{figure}
\centering
\includegraphics[width = 3cm]{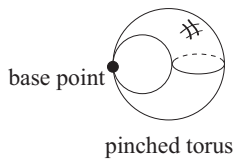}
\caption{A pinched torus, which is obtained by shrinking one nontrivial loop on a torus to a point. The domain of $K_2 (\mathcal{M}, \phi_0)$ is homeomorphic to the pinched torus.
}\label{fig-8}
\end{figure}

\begin{figure}
\centering
\includegraphics[width = 10cm]{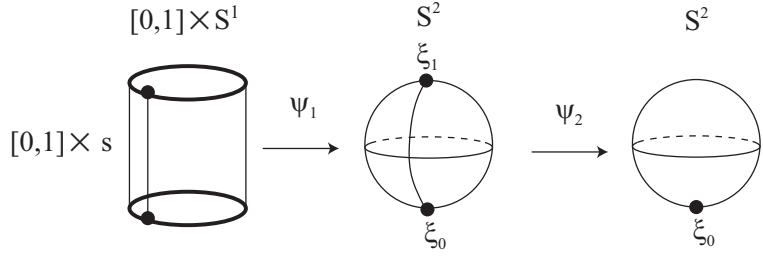}
\caption{Successive deformations of $[0,1] \times S^{1}$ by $\psi_1$ and $\psi_2$. The two loops $0\times S^1$ and $1\times S^1$ are mapped by $\psi_1$ to two dots, which are then mapped by $\psi_2$ to a single point $\xi_0$. In the Abe homotopy group, the space mapped to the base point $\phi_0$ is $0 \times S^1 \cup 1 \times S^1$, which is shown by the bold loops and dots. Hence, the space mapped to $\phi_0$ is different from that in Fig. \ref{fig-2}.}\label{fig-8-b}
\end{figure}

 For the case of $n=2$, $K_2 (\mathcal{M}, \phi_0)$ is a set of the maps from $([0,1] \times S^1 , 0 \times S^1 \cup 1 \times S^1 )$ to $(\mathcal{M},\phi_0)$. If we identify all points on the top loop $ 1 \times S^1$ and the bottom loop $  0 \times S^1$, the domain space $[0,1] \times S^1$ becomes a pinched torus as illustrated in Fig. \ref{fig-8}.
The $n$th Abe homotopy group is shown to be isomorphic to a semi-direct product of $\pi_1 (\mathcal{M}, \phi_0)$ and $\pi_n (\mathcal{M}, \phi_0)$ by Abe \cite{Abe;1940}. Here, we review his proof.
First, we show that $\pi_n (\mathcal{M}, \phi_0)$ is isomorphic to a subgroup of $\kappa_n (\mathcal{M}, \phi_0)$. The map $f^{\rm a}_{\phi_0}$ is a composition of $\psi_1$ and ${f_{\phi_0}^{\rm a}}'$ that maps $(S^n,\xi_0 \cup \xi_1)$ to $(\mathcal{M}, \phi_0)$. Because $\psi_1$ only changes  the space mapped to $\phi_0 $ as shown in Fig.~\ref{fig-8-b}, $f_{\phi_0}^{\rm a}$ is isomorphic to ${f_{\phi_0}^{\rm a}}'$:
\begin{equation}
 f_{\phi_0}^{\rm a} = {f_{\phi_0}^{\rm a}}' \circ \psi_1. \label{A2-1-3}
\end{equation}
We define a map $f_{\phi_0}$ such that
\begin{equation}
 f_{\phi_0} : (S^n , \xi_0 ) \to (\mathcal{M}, \phi_0). \label{A2-1-4}
\end{equation}
Then, $f_{\phi_0}$ is related to ${f^{\text{a}}_{\phi_0}}'$ by $\psi_2$,
\begin{equation}
{f_{\phi_0}^{\rm a}}' = f_{\phi_0} \circ \psi_2 . \label{A2-1-5}
\end{equation}
Note here that this is different from Eq. (\ref{A2-1-2}). This map is not isomorphism since we can arbitrarily choose path $\overline{\xi_0\xi_1}$ which is mapped to $\xi_0$ by $\psi_2$. By Eqs. (\ref{A2-1-3}) and (\ref{A2-1-5}), we define a map from $(\mathcal{M}, \phi_0)^{(S^n, \xi_0)}$ to $K_n (\mathcal{M}, \phi_0)$ as
\begin{equation}
 \Psi_1 : f_{\phi_0}(t,x) \mapsto f^{\rm a}_{\phi_0} (t,x) = f_{\phi_0} \circ \psi_2 \circ \psi_1(t,x) \;  \text{for} \;  ^{\forall} t \in [0,1], ^{\forall} x \in S^{n-1}. \label{A2-1-6}
\end{equation}
The homotopy equivalence class of $f_{\phi_0}^{\rm a}$ corresponds to an element of $\kappa_n (\mathcal{M},\phi_0)$, whereas that of $f_{\phi_0}$ corresponds to an element of $\pi_n (\mathcal{M}, \phi_0)$. Therefore, $\psi_2 \circ \psi_1 $ induces a map ${\Psi_1}_{\ast}$ from $\pi_n (\mathcal{M},\phi_0)$ to $\kappa_n (\mathcal{M}, \phi_0)$ defined by
\begin{equation}
\begin{split}
 {\Psi_1}_{\ast} :& \pi_n (\mathcal{M}, \phi_0 ) \to \kappa_n (\mathcal{M}, \phi_0 ), \\ 
&[f_{\phi_0}(t,x)] \mapsto [f_{\phi_0} \circ \psi_2 \circ \psi_1 (t,x)] \;  \text{for} \;  ^{\forall} t \in [0,1], ^{\forall} x \in S^{n-1}. \label{A2-1-7}
\end{split}
\end{equation}
The map ${\Psi_1}_{\ast}$ is a homomorphic map.
From the general property of homomorphism, $\pi_n (\mathcal{M}, \phi_0)$ is isomorphic to a subgroup of $\kappa_n (\mathcal{M}, \phi_0)$ if $\Psi_1$ is injective. Actually, ${\Psi_1}_{\ast}$ is proved to be injective, since when $[f_{\phi_0}''] \in \ker {\Psi_1}_{\ast}$, i.e., $f_{\phi_0}'' \circ \psi_2 \circ \psi_1 \sim 1_{\phi_0}$, we immediately obtain $f''_{\phi_0} \sim 1_{\phi_0}$. Here $1_{\phi_0} $ is the map to $\phi_0$. Hence, $\pi_n (\mathcal{M}, \phi_0)$ is isomorphic to a subgroup of $\kappa_n (\mathcal{M}, \phi_0)$. 	

Next, we show that $\kappa_n (\mathcal{M}, \phi_0)$, $\pi_n (\mathcal{M}, \phi_0)$, and $\pi_1 (\mathcal{M}, \phi_0)$ satisfy the following relation:
\begin{equation}
 \kappa_n(\mathcal{M}, \phi_0) / \pi_n(\mathcal{M}, \phi_0 ) \cong \pi_1 (\mathcal{M}, \phi_0). \label{A2-1-8}
\end{equation}
 We restrict a domain of $f_{\phi_0}^{\rm a} (t,x)$ to $[0,1] \times s$, where $s$ is a fixed point in $S^{n-1}$. Since $f^{\rm a}_{\phi_0} ( t,s) $ satisfies $f_{\phi_0}^{\rm a}(0,s) = f^{\rm a}_{\phi}(1,s) = \phi_0$, $f_{\phi_0}^{\rm a}$ forms a loop on $\mathcal{M}$ with base point $\phi_0$. Thus, the homotopy equivalence class of $f_{\phi_0}^{\rm a} (t,s)$ corresponds to an element of $\pi_1 (\mathcal{M}, \phi_0)$. We identify the map $l_{\phi_0} (t) \in (\mathcal{M}, \phi_0)^{([0,1], 0 \cup  1)}$ such that $l_{\phi_0} (t) = f^{\rm a}_{\phi_0 } (t,s)$ for any $t$, and define $\Psi_2$ as a projective map from $K_n (\mathcal{M}, \phi_0)$ to $ (\mathcal{M}, \phi_0)^{([0,1], 0 \cup1)}$:
\begin{equation}
 \Psi_2 : f_{\phi_0}^{\text{a}}(t,x) \mapsto l_{\phi_0} (t) = f^{\rm a}_{\phi_0} (t, s) \;  \text{for} \;  ^{\forall} t \in [0,1], ^{\forall} x \in S^{n-1}. \label{A2-1-9}
\end{equation}
Here, $\Psi_2$ can naturally induce surjective homomorphism ${\Psi_2}_{\ast}$ from $\kappa_n (\mathcal{M}, \phi_0)$ to $\pi_1 (\mathcal{M}, \phi_0)$:
\begin{equation}
\begin{split}
 {\Psi_2}_{\ast} : & \kappa_n (\mathcal{M}, \phi_0) \to \pi_1 (\mathcal{M}, \phi_0) ,\\
& [f_{\phi_0}^{\rm a}(t,x)] \mapsto [ l_{\phi_0} (t)]= [f^{\rm a }_{\phi_0} (t, s)]  \;  \text{for} \; ^{\forall} t \in [0,1],  ^{\forall} x \in S^{n-1} . \label{A2-1-10}
\end{split}
\end{equation}
Moreover, $\ker {\Psi_2}_{\ast}$ is shown to be isomorphic to $\pi_n (\mathcal{M}, \phi_0)$: for any $[{f_{\phi_0}^{\rm a}}''] (t,x) \in \ker {\Psi_2}_{\ast} $, ${f_{\phi_0}^{\rm a}}'' (t,x)$ satisfies ${f_{\phi_0}^{\rm a}}''(t,s)=\phi_0$ for any $t \in [0,1]$. Then, ${f_{\phi_0}^{\rm a}}''(t,x)$ satisfies Eq. (\ref{A2-1-1}) and the homotopy equivalence class of ${f_{\phi_0}^{\rm a}}''(t,x)$ becomes an element of $\pi_n (\mathcal{M}, \phi_0)$. Therefore, from the homomorphism theorem for ${\Psi_2}_{\ast}$, we obtain Eq. (\ref{A2-1-8}). Note that, for given $\gamma \in \pi_1 (\mathcal{M}, \phi_0)$ and $\alpha \in \pi_n (\mathcal{M}, \phi_0)$, we obtain $\gamma^{-1} \ast \alpha \ast \gamma \in \pi_n (\mathcal{M}, \phi_0)$, and hence, $\pi_n (\mathcal{M},\phi_0)$ is a normal subgroup of $\kappa_n (\mathcal{M}, \phi_0)$. 
 
Let us define an inclusion map $\iota$ from $(\mathcal{M}, \phi_0)^{([0,1], 0 \cup 1)}$ to $(\mathcal{M}, \phi_0)^{( [0,1] \times s,0\times s \cup  1  \times s)} $ such that
\begin{equation}
\iota : l_{\phi_0} (t) \mapsto f^{\rm a}_{\phi_0} (t, s) = l_{\phi_0} (t)  \;  \text{for} \;^{\forall}t \in [0,1] , s \in S^{n-1}. \label{A2-1-11}
\end{equation} 
This induces a map from $\pi_1 (\mathcal{M}, \phi_0)$ to $\kappa_n (\mathcal{M}, \phi_0)$ as
\begin{equation}
\begin{split}
 \iota_{\ast} :& \pi_1 (\mathcal{M}, \phi_0 ) \to \kappa_n (\mathcal{M}, \phi_0 ) \\
   &[ l_{\phi_0} (t)] \mapsto [f^{\rm a}_{\phi_0} (t, s)] = [l_{\phi_0}(t)]  \;  \text{for} \; ^{\forall}t \in [0,1] , s \in S^{n-1}. \label{A2-1-12}
\end{split}
\end{equation}
Here, $\iota_{\ast}$ satisfies $(\Psi_2)_{\ast} \circ \iota_{\ast} = \id_{\pi_1} $, where $\id_{\pi_1}$ is an identity map in $\pi_1 (\mathcal{M}, \phi_0)$. We note that ${\Psi_1}_{\ast}$, ${\Psi_2}_{\ast}$, and $\iota_{\ast}$ constitute the following diagram:
\def\llongleftarrow{\leftarrow\joinrel\relbar\joinrel\relbar\joinrel\relbar}
\def\llongrightarrow{\relbar\joinrel\relbar\joinrel\relbar\joinrel\rightarrow}
\def\psimap{\stackrel{{\Psi_2}_{\ast}}{\llongrightarrow}}
\def\imap{\stackrel{\llongleftarrow}{\iota_{\ast}}}
\def\equiarrow{\mathrel{\lower.85ex\hbox{$\stackrel{\psimap}{\imap} $}}}
\begin{equation}
 \pi_n (\mathcal{M}, \phi_0 ) \xrightarrow{{\Psi_1}_{\ast}} \kappa_n (\mathcal{M}, \phi_0) \equiarrow  \pi_1 (\mathcal{M}, \phi_0). \label{A-2-8}
\end{equation}
which implies that $\kappa_n (\mathcal{M}, \phi_0)$ is a semi-direct product of $\pi_1 (\mathcal{M}, \phi_0)$ and $\pi_n (\mathcal{M}, \phi_0)$ because ${\Psi_1}_{\ast}$ is injective, ${\Psi_2}_{\ast} $ is surjective, ${\Psi_1}_{\ast}$ and ${\Psi_2}_{\ast}$ satisfy $\ker {\Psi_2}_{\ast} \cong \pi_n (\mathcal{M}, \phi_0) \cong \im {\Psi_1}_{\ast}$, and elements of $\pi_1$ and those of $\pi_n$ do not commute in general.
Therefore, the following theorem holds \cite{Abe;1940,Fox;1948}. 

\begin{thm}
The $n$th Abe homotopy group $\kappa_n $ is isomorphic to the semi-direct product of $\pi_1 (\mathcal{M}, \phi_0)$ and $\pi_n (\mathcal{M}, \phi_0)${\rm :} 
 \begin{equation}
  \kappa_n (\mathcal{M}, \phi_0) \cong \pi_1(\mathcal{M},\phi_0) \ltimes \pi_n (\mathcal{M}, \phi_0). \label{A2-4}
 \end{equation}
Here, an element of $\kappa_n (\mathcal{M}, \phi_0)$ is expressed as a set of elements of $\pi_n (\mathcal{M}, \phi_0)$ and $\pi_1 (\mathcal{M}, \phi_0)$, and for any $\gamma_1, \gamma_2 \in \pi_1 (\mathcal{M}, \phi_0)$ and $\alpha_1 , \alpha_2 \in \pi_n (\mathcal{M}, \phi_0)$, the semi-direct product is defined by
 \begin{equation}
  (\gamma_1 ,\alpha_1 ) \ast (\gamma_2 , \alpha_2) = (\gamma_1 \ast \gamma_2, \gamma_2 (\alpha_1) \ast \alpha_2 ), \label{A2-5}
 \end{equation}
  where $\gamma (\alpha) = \gamma^{-1} \ast \alpha \ast \gamma$. 
\end{thm} \label{thm-1}

\subsection{Abe homotopy group and topological excitations}
\label{sec:infl-Abe}
In this section, we show that the charge of a topological excitation, which coexists with vortices, is defined by the conjugacy class of the Abe homotopy group. 

\begin{figure}
\centering
\includegraphics[width=8cm]{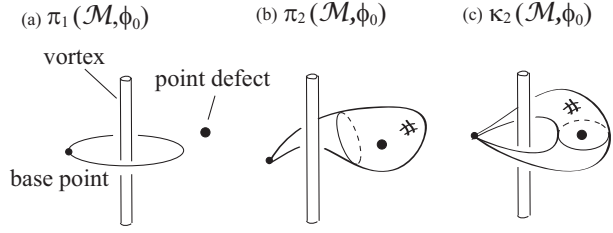}
\caption{Domains of elements of (a) $\pi_1 (\mathcal{M},\phi_0)$, (b) $\pi_2 (\mathcal{M},\phi_0)$, and (c) $\kappa_2(\mathcal{M}, \phi_0)$. (a) The domain of $\pi_1 (\mathcal{M}, \phi_0)$ is a loop encircling a vortex. (b) The domain of $\pi_2 (\mathcal{M}, \phi_0)$ is a closed surface enclosing a point defect. (c) The domain of $\kappa_2 (\mathcal{M}, \phi_0)$ is a pinched torus encircling a vortex and enclosing a point defect simultaneously. 
} \label{fig-5}
\end{figure}

 By Theorem 1, an element of the $n$th Abe homotopy group $\kappa_n (\mathcal{M}, \phi_0)$ is described by a set of elements $\gamma \in \pi_1 (\mathcal{M}, \phi_0)$ and $\alpha \in \pi_n (\mathcal{M}, \phi_0)$, which means that $\kappa_n (\mathcal{M}, \phi_0)$ can simultaneously classify vortices and topological excitations with the dimension of homotopy $n \ge 2 $ as shown in Fig. \ref{fig-5}. When $\gamma = 1_{\phi_0}$, $(1_{\phi_0}, \alpha)$ represents a situation in which there is only a topological excitation with $n \ge 2$. On the other hand, $\alpha = 1_{\phi_0}$ implies $(\gamma, 1_{\phi_0})$, which indicates that there is only a vortex. 

 We are interested in the case in which there are both vortices and topological excitations with $n \ge 2$. Let us consider a coalescence of $(\gamma_1,\alpha)$ and $(\gamma_2,1_{\phi_0})$: 
\begin{equation}
 (\gamma_1 , \alpha ) \ast (\gamma_2,1_{\phi_0}) = (\gamma_1 \ast \gamma_2, \gamma_2(\alpha) ). \label{S3-8}
\end{equation}
Note that $\gamma_2$ acts not only on $\gamma_1$ but also on $\alpha$. As discussed in Sec. \ref{sec:homotopy-exp}, $\gamma_2 (\alpha)$ represents topological excitations after going through a closed loop $\gamma_2$. Therefore, the effect of vortices on topological excitations is included in the algebra of the $n$th Abe homotopy group. Note that the influence of $\gamma$ on $\alpha$ originates from noncommutativity between $\pi_1 (\mathcal{M}, \phi_0)$ and $\pi_n (\mathcal{M}, \phi_0)$: if $\gamma$ and $\alpha$ are noncommutative, there is an influence since $\gamma (\alpha ) \neq \alpha$, whereas if they are commutative, there is no influence. For the latter case, Eq. (\ref{A2-5}) can be rewritten as
\begin{equation}
 (\gamma_1 , \alpha_1) \ast (\gamma_2 ,\alpha_2 ) = (\gamma_1 \ast \gamma_2 , \alpha_1 \ast \alpha_2 ), \label{S3-9-2}
\end{equation}
which means the Abe homotopy group is given by the direct product of $\pi_1(\mathcal{M}, \phi_0)$ and $\pi_n (\mathcal{M}, \phi_0)$.

Exchanging the order of the product in Eq. (\ref{S3-8}), we obtain
\begin{equation}
 (\gamma_2,1_{\phi_0}) \ast (\gamma_1 , \alpha ) = (\gamma_2 \ast \gamma_1 , \alpha ).  \label{S3-10}
\end{equation}
In this case, $\gamma_2$ does not act on $\alpha$ and thus there is no influence of the vortex. From this point of view, elements of $\kappa_n (\mathcal{M}, \phi_0)$ are noncommutative. Higher-dimensional topological excitations characterized by $\kappa_n (\mathcal{M}, \phi_0)$ depends on the choice of the path which passes by vortices in a manner similar to the case of $\pi_1 (\mathcal{M}, \phi_0)$ (see Sec. \ref{sec:homotopy}). To define a charge which is independent of the choice of the path, we take the conjugacy classes of $\kappa_n (\mathcal{M}, \phi_0)$ as in the case of $\pi_1 (\mathcal{M}, \phi_0)$. In other words, we construct classes of $\kappa_n (\mathcal{M}, \phi_0)$ that are invariant under inner automorphism. Since all elements of $\pi_n (\mathcal{M}, \phi_0)$ commute with each other, we only consider the noncommutativity of elements of $\pi_1 (\mathcal{M}, \phi_0)$ and that of elements of $\pi_1 (\mathcal{M}, \phi_0)$ and $\pi_n (\mathcal{M}, \phi_0)$, and derive the equivalence relation between them.

\begin{figure}[tbp]
 \centering
 \includegraphics[width=10cm]{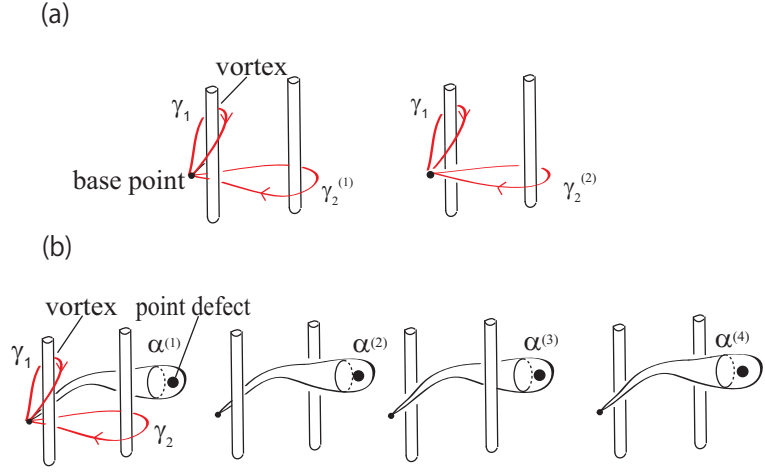}
 \caption{(Color online)
(a) Two possible configurations of two vortices which are represented by loops $\gamma_1$ and $\gamma_2$. For $\gamma_2$, there are two homotopically inequivalent paths: one goes behind the vortex $\gamma_1$ and the other goes in front of it. The former is labeled $\gamma_2^{(1)}$ and the latter is labeled $\gamma_2^{(2)}$. They are related by $\gamma^{-1}_1 \ast \gamma_2^{(1)}\ast \gamma_1 = \gamma_2^{(2)}$, as illustrated Fig. \ref{fig-1}. (b) Four possible configurations of two vortices labeled $\gamma_1$ and $\gamma_2$ and a point defect which is represented by a closed surface labeled $\alpha$. They satisfy the relations (\ref{S3-12-a}) - (\ref{S3-12-d}). 
}\label{fig-6}
\end{figure}

In Fig. \ref{fig-6}, we show possible configurations of paths describing the elements of $\pi_1$ and $\pi_n$. For any $\gamma_1, \gamma_2^{(i)} \in \pi_1 (\mathcal{M}, \phi_0)\; (i=1,2)$ in Fig. \ref{fig-6}(a), we obtain the relation:
\begin{equation}
  \gamma_1^{-1} \ast \gamma_2^{(1)} \ast \gamma_1 = \gamma_2^{(2)}. \label{S3-11}
\end{equation}
For any $\gamma_1 ,\gamma_2 \in \pi_1 (\mathcal{M}, \phi_0)$ and $\alpha^{(j)} \in \pi_n (\mathcal{M}, \phi_0) \; (j=1,2,3,4)$ in Fig. \ref{fig-6}(b), the following relations hold: 
\begin{subequations}
\begin{align}
& \gamma_1^{-1} \ast \alpha^{(1)} \ast \gamma_1 = \alpha^{(3)},  \label{S3-12-a} \\
& \gamma_2^{-1} \ast \alpha^{(1)} \ast \gamma_2 = \alpha^{(2)},  \label{S3-12-b} \\
& \gamma_1 \ast \alpha^{(4)} \ast \gamma_1^{-1} = \alpha^{(2)},  \label{S3-12-c} \\
& \gamma_2 \ast \alpha^{(4)} \ast \gamma_2^{-1}  = \alpha^{(3)}, \label{S3-12-d} \\
& \gamma_2 \ast \gamma_1 \ast \alpha^{(4)} \ast \gamma_1^{-1} \ast \gamma_2^{-1} = \alpha^{(1)},  \label{S3-12-e} \\
& \gamma_2^{-1} \ast \gamma_1 \ast \alpha^{(3)} \ast \gamma_1^{-1} \ast \gamma_2 = \alpha^{(2)}. \label{S3-12-f}
\end{align}
\end{subequations}
Equations (\ref{S3-12-a})-(\ref{S3-12-d}) give the complete set of relations among $\alpha^{(i)} \; (i=1,2, 3,4)$,  and Eqs. (\ref{S3-12-e}) and (\ref{S3-12-f}) are derived from Eqs. (\ref{S3-12-a}) - (\ref{S3-12-d}).

We define a topological charge so that it does not depend on the choice of the paths. For vortices, we impose the equivalence relation $\gamma_2^{(1)} \sim \gamma_2^{(2)}$ on Eq.~(\ref{S3-11}):
\begin{equation}
 \gamma_1^{-1} \ast \gamma_2^{(i)} \ast \gamma_1 \sim \gamma_2^{(i)} \; \text{for} \;  ^{\forall}\gamma_1 \in \pi_1 (\mathcal{M} , \phi_0 ), \label{S3-13}
\end{equation}
where $i=1,2$. For higher-dimensional topological excitations, we also impose the equivalence relation $\alpha^{(i)} \sim \alpha^{(j)} \ \ (i,j = 1,2,3,4)$ on Eqs. (\ref{S3-12-a})-(\ref{S3-12-d}): 
\begin{equation}
  \gamma_1^{-1} \ast \alpha^{(j)} \ast \gamma_1 \sim \alpha^{(j)} \; \text{for}\;  ^{\forall}\gamma_1 \in \pi_1 (\mathcal{M} , \phi_0 ). \label{S3-14}
\end{equation}
Therefore, the charge of the topological excitation that coexists with vortices characterizes the conjugacy classes of the Abe homotopy group.
 Relation (\ref{S3-14}) is equivalent to the equivalence relation (\ref{S3-5}) for $\pi_n (\mathcal{M}, \gamma_1)$. However, for the case of $\pi_n (\mathcal{M}, \gamma_1)$, the element of $\pi_1 (\mathcal{M}, \phi_0)$ is fixed.

From (\ref{S3-13}) and (\ref{S3-14}), we have 
\begin{equation}
\begin{split}
 (1_{\phi_0} , \alpha_1) \ast (1_{\phi_0} , \alpha_2) &= (\gamma \ast \gamma^{-1} , \alpha_1 ) \ast (1_{\phi_0} , \alpha_2) \\
                                                                        &\sim (\gamma , \alpha_1 ) \ast (\gamma^{-1}, 1_{\phi_0} ) \ast (1_{\phi_0} , \alpha_2) \\
                                                                       &=  (\gamma , \alpha_1) \ast (\gamma^{-1} ,\alpha_2 ) \\
                                                                          & = (1_{\phi_0},\gamma^{-1}(\alpha_1) \ast \alpha_2). \label{S3-16}
\end{split}
\end{equation}
This result indicates that even if there is no vortex in the initial state, an influence of vortices emerges through vortex-anti-vortex pair creation.

We summarize the relations between $n$-simple, noncommutativity of $\pi_1$ and $\pi_n$, and the semi-direct product in the Abe homotopy group.
\begin{thm}
The following four statements are equivalent with each other{\rm :}
 \begin{itemize}
   \item[{\rm (a)}] The influence of vortices exists. 
   \item[{\rm (b)}] The order parameter manifold is not $n$-simple.
   \item[{\rm (c)}] Elements of $\pi_1$ and those of $\pi_n$ do not commute with each other. 
   \item[{\rm (d)}] The $n$th Abe homotopy group $\kappa_n (\mathcal{M}, \phi_0)$ is isomorphic to the nontrivial semi-direct product of $\pi_1(\mathcal{M},\phi_0)$ and $\pi_n (\mathcal{M}, \phi_0)$. 
 \end{itemize}
\end{thm}

From Theorem 2, we can check the influence of vortices and the structure of $n$th Abe homotopy group by calculating $\gamma (\alpha)= \gamma^{-1} \ast \alpha \ast \gamma$.

\section{Calculation of the influence}
\label{sec:calc}
In this section, we calculate $\gamma(\alpha)$ by making use of Eilenberg's theory \cite{Eilenberg;1939}. We also use the theory of orientation of manifolds (see Refs.~\cite{Greenberg:1981}) to calculate the influence of $\pi_1$ on $\pi_n$. The goal of this section is to derive Theorem 4 and Corollary 1 which give simple rules for the calculation of $\gamma (\alpha)$.
\subsection{The case with $\mathcal{M} \simeq S^n/K$}
\label{sec:calc-1}
We consider the case in which the order parameter manifold is given by
\begin{equation}
 \mathcal{M} \simeq SO(n+1)/(SO(n) \rtimes K) \simeq S^n /K \ \ (n \ge 2),
 \end{equation}
where $S^n$ is an $n$-dimensional sphere embedded in $\mathbb{R}^{n+1}$ and $K$ is a discrete subgroup of $SO(n+1)$, which freely acts on $S^n$. Hence, the nontrivial element $ g \in K$ has no fixed points in $S^n$. For any $x \in S^n$ and $g \in K $, $S^n/K$ is defined as a set of equivalence classes under the equivalence relation $x \sim g(x)$. Here, we focus on the case of $\mathcal{M} \simeq S^n/K$, since the results for this case are applicable for many physical systems. Examples will be given in Sec. \ref{sec:app}. The homotopy groups of $S^n/K$ are given by  
\begin{subequations}
 \begin{align}
  \pi_1 (S^n /K, x_0 ) &\cong K, \label{S3-2-2-a}\\
  \pi_k (S^n/K, x_0 ) & \cong 0  \ \ (1 < k < n), \label{S3-2-2-b}\\
  \pi_n (S^n/K, x_0) & \cong \mathbb{Z},  \label{S3-2-2-c}
 \end{align}
\end{subequations}
where $x_0$ is a base point on $S^n/K$.

\begin{figure}[tbtp]
 \centering
 \includegraphics[width=5cm]{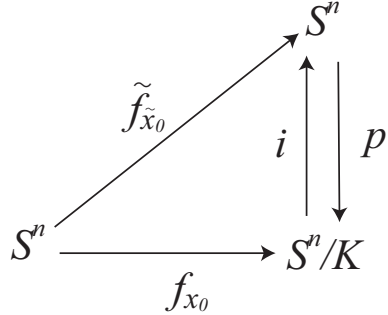}
 \caption{
Diagram showing the actions of $f_{x_0} : (S^n,s) \to (S^n/K,x_0)$, inclusion map $i :  (S^n/K,x_0) \to (S^n,\tilde{x}_0) $, projection $p$ from $S^n$ to $S^n/K$, and composite map $\tilde{f}_{\tilde{x}_0} := i \circ f_{x_0} : S^n  \to S^n$ between $i$ and $f_{x_0}$. 
}\label{fig:map}
\end{figure}

\begin{figure}[tbp]
 \centering
 \includegraphics[width=3cm]{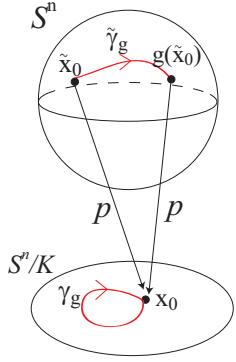}
 \caption{(Color online)
Lifting from $(S^n/K,x_0)$ to $(S^n,\tilde{x}_0)$. A loop $\gamma_g$ with a base point $x_0$ on $S^n/K$ is mapped by $i$ onto path $\tilde{\gamma}_g$ that connects $\tilde{x}_0$ to $g(\tilde{x}_0)$ on $S^n$ under the lift, where $p(\tilde{x}_0) = p (g(\tilde{x}_0)) = x_0$.
} \label{fig-7}
\end{figure}

 A universal covering space of $S^n/K$ is $S^n$ and there is a projection map $p:S^n \to S^n/K$. An inverse of $p$ is given by an inclusion map $i \ \  (p \circ i= \id_{S^n/K})$, which is a map from $S^n/K$ to $S^n$. We call this map $i$  {\it lift} in this paper.
Define a map $f_{x_0}$ with a base point $x_0 \in S^n/K$ as
\begin{equation}
f_{x_0} : (S^n,s)  \to (S^n/K,x_0),
\end{equation}
where $s$ is a point on the domain space $S^n$. By the homotopy lifting lemma (see e.g., Section 5 in Ref.~\cite{Greenberg:1981}), we can uniquely specify the mapping from $(S^n,s)$ to the universal covering space of $(S^n/K,x_0)$ under the lift as
\begin{equation} 
\tilde{f}_{\tilde{x}_0} : (S^n,s) \to (S^n,\tilde{x}_0),
\end{equation}
where $\tilde{f}_{\tilde{x}_0}$ is defined by $i$ as $\tilde{f}_{\tilde{x}_0} = i \circ f_{x_0}$. Here, $\tilde{x}_0 $ is a base point on the universal covering space $S^n$ such that $p (\tilde{x}_0) =x_0$. The relations among $f_{x_0}$, $\tilde{f}_{\tilde{x}_0}$, $p$, and $i$ are shown in Fig. \ref{fig:map}.
Similarly, a map from $(S^1,s)$ to $(S^n/K,x_0)$ is lifted by $i$. Let $\gamma_g$ be a nontrivial element of $\pi_1(S^n/K,x_0)$ corresponding to $g \in K$. As shown in Fig. \ref{fig-7}, $\gamma_g$ is a closed path on $S^n/K$, while its lift $\tilde{\gamma}_g$ is a path connecting $\tilde{x}_0$ and $g (\tilde{x}_0)$. In the calculation of $\gamma_g (\alpha)$, we first consider a map $f_{x_0}$ such that $\alpha = [f_{x_0}]$, and then investigate a map $\gamma_g (f_{x_0})$. Since the homotopy lifting lemma also holds for $\gamma_g (f_{x_0})$, the lifted map $\tilde{\gamma}_g (\tilde{f}_{\tilde{x}_0})$ is uniquely determined. In what follows, we calculate the degree of mapping of $\tilde{\gamma} (\tilde{f}_{\tilde{x}_0})$ which has a one-to-one correspondence with the homotopy equivalence classes of $\gamma (f_{x_0})$.  

An element of $\pi_n (S^n,\tilde{x}_0)$ has a one-to-one correspondence with a degree of mapping from $S^n$ to $S^n$. A degree of mapping is defined as follows: let $\omega$ be an $n$-form on $S^n$ and $m(\tilde{x})$ be a metric tensor on $\tilde{x}$, where $\tilde{x}$ is an arbitrary point on the universal covering space $S^n$ of $S^n/K$. Then, a volume form is defined by  
\begin{equation}
 \omega = \sqrt{\det (m(\tilde{x}))} d\tilde{x}^1 \wedge d\tilde{x}^2 \wedge \cdots \wedge d\tilde{x}^n, \label{eil-4}
\end{equation}
where $(\tilde{x}^1,\tilde{x}^2, \cdots , \tilde{x}^n)$ is a coordinate system on $S^n$. Let $(\theta^1(\tilde{x}), \theta^2(\tilde{x}), \cdots ,\theta^n(\tilde{x}))$ be an orthogonal basis at $\tilde{x} \in S^n$. Then, the volume form is rewritten as
\begin{equation}
 \omega  = \theta^1(\tilde{x}) \wedge \theta^2(\tilde{x}) \wedge \cdots \wedge \theta^n(\tilde{x}). \label{eil-2}
\end{equation}
By using $\omega$ in Eq. (\ref{eil-4}), a degree of mapping is defined by
\begin{equation} 
 \deg (\tilde{f}_{\tilde{x}_0}) : = \int_{S^n} \tilde{f}_{\tilde{x}_0}^{\ast} \omega, \label{eil-2-2}
\end{equation}
where $\tilde{f}_{\tilde{x}_0}^{\ast}$ is the pullback map induced by $\tilde{f}_{\tilde{x}_0}$. 
Next, we calculate $\deg(\tilde{\gamma}_g(\tilde{f}))$ by using the relation derived by Eilenberg \cite{Eilenberg;1939,Eilenberg:1939th}:
\begin{equation}
  (\tilde{\gamma}_g (\tilde{f}_{\tilde{x}_0}))^{\ast} \omega =  \tilde{f}_{\tilde{x}_0}^{\ast} \circ g^{\ast} \omega + d \chi, \label{eil-3}
\end{equation}
where $g^{\ast}$ is the pullback of $g$, where $g \in K$ is an action on the order parameter space $ g  : (S^n, \tilde{x}_0) \mapsto (S^n ,g(\tilde{x}_0))$. Here, we have rewritten the relation in differential forms, whereas the original one derived in Ref.~\cite{Eilenberg;1939} was written in terms of the homology group. Since Eq. (\ref{eil-3}) is defined for the de Rham cohomology group, this transformation involves an arbitrary term $d \chi$. In Eq. (\ref{eil-3}), the influence of $\tilde{\gamma}_g$ on $\tilde{f}_{\tilde{x}_0}$ is translated into the action of $g \in K$ on the manifold $S^n$.  

\begin{figure}[tbp]
 \centering
 \includegraphics[width=5cm]{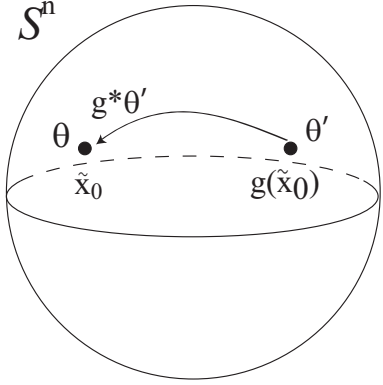}
 \caption{
$\theta^A \ \ (A=1,2,\cdots ,n)$ is an orthogonal basis set on $\tilde{x}_0 \in S^n$ and $\theta'^{A}$ is another on $g(\tilde{x}_0) \in S^n$. $g^{\ast}$ is the pullback map of $g \in K$ that maps $\theta'^{A}$ to $g^{\ast}\theta'^{A}$ at  $\tilde{x}_0$. Since both $\theta^A$ and $g^{\ast} \theta'^A$ are defined at $\tilde{x}_0$, they transform into each other by an orthogonal transformation. 
}\label{fig-9}
\end{figure}

We derive a relationship between $g^{\ast} \omega$ and $\omega$.
Defining $\tilde{x}'=g(\tilde{x}) \ \ (^{\forall}\tilde{x} \in S^n)$, the volume form on $\tilde{x}'$ is given by
\begin{equation}
\begin{split}
 \omega'(\tilde{x}') & = \omega (g(\tilde{x})) \\
                            &= \sqrt{\det( m(\tilde{x}'))} d\tilde{x}'^1 \wedge d\tilde{x}'^2 \wedge \cdots \wedge d\tilde{x}'^n. \label{eil-15}
\end{split}
\end{equation}
By using the orthogonal basis, Eq. (\ref{eil-15}) is rewritten as
\begin{equation}
 \omega'(\tilde{x}') = \theta'^1(\tilde{x}') \wedge \theta'^2(\tilde{x}') \wedge \cdots \wedge \theta'^n(\tilde{x}'). \label{eil-5}
\end{equation}
The pullback map $g^{\ast} $ acts on $\omega'$ as
\begin{equation}
  (g^{\ast} \omega') (\tilde{x})=  g^{\ast}\theta'^1 \wedge g^{\ast}\theta'^2 \wedge \cdots \wedge g^{\ast}\theta'^n. \label{eil-6}
\end{equation}
The orthogonal bases $(g^{\ast}\theta'^1 , g^{\ast}\theta'^2 , \cdots , g^{\ast}\theta'^n)$ and $(\theta^{1}, \theta^2, \cdots, \theta^n )$ are defined on the same point $\tilde{x} \in S^n$ as shown in Fig. \ref{fig-9}, and therefore there is an orthogonal transformation $O \in O(n)$ such that
\begin{equation}
g^{\ast} \theta'^{A}= O^A_{\ \ B} \theta^B, \label{eil-7}
\end{equation}
where $A,B =1,2,\cdots ,n$. The relation between $g^{\ast} \omega'$ and $\omega$ is given by using $O \in O(n)$ as
\begin{equation}
 g^{\ast} \omega' = \det(O) \omega. \label{eil-8}
\end{equation}

By definition (\ref{eil-2-2}), the degree of mapping of $\tilde{\gamma}_g(\tilde{f}_{\tilde{x}_0})$ is calculated as follows:
\begin{equation}
\begin{split}
 \deg( \tilde{\gamma}_g(\tilde{f}_{\tilde{x}_0})) &= \int_{S^n} (\tilde{\gamma}_g(\tilde{f}_{\tilde{x}_0}))^{\ast} \omega \\
                                                   &= \int_{S^n} (\tilde{f}_{\tilde{x}_0}^{\ast} \circ g^{\ast} \omega + d \chi ) \\
                                                   &= \int_{S^n} \tilde{f}_{\tilde{x}_0}^{\ast} \circ g^{\ast} \omega+  \int_{S^n} d \chi \\
                                                    &= \det(O) \int_{S^n} \tilde{f}_{\tilde{x}_0}^{\ast} \omega \\
                                                    &= \det(O) \deg(\tilde{f}_{\tilde{x}_0}) \label{eil-5-2}
\end{split}
\end{equation}
 Consequently $\deg(\tilde{f}_{\tilde{x}_0})$ and $\deg (\tilde{\gamma}_g(\tilde{f}_{\tilde{x}_0}))$ differ only by a factor of $\det (O) = \pm 1$, where the sign of $\det (O)$ depends only on $g \in K$. Defining $\deg (\tilde{\gamma}_g) := \det (O)$, we can rewrite $ \deg(\tilde{\gamma}_g (\tilde{f}_{\tilde{x}_0}) )$ as 
\begin{equation}
 \deg(\tilde{\gamma}_g (\tilde{f}_{\tilde{x}_0}) ) = \deg(\tilde{\gamma}_g) \deg(\tilde{f}_{\tilde{x}_0}). \label{eil-9}
\end{equation}
Since there is a one-to-one correspondence between $[\tilde{f}_{\tilde{x}_0}] \in \pi_n (S^n,\tilde{x}_0)$ and $\deg(\tilde{f}_{\tilde{x}_0})$ through the Hurewise map and de Rham duality \cite{Eilenberg:1939th}, we obtain
\begin{equation}
 [\tilde{\gamma}_g (\tilde{f}_{\tilde{x}_0})] = [\tilde{f}_{\tilde{x}_0}]^{\deg(\tilde{\gamma}_g)},\label{eil-10}
\end{equation}
where $\deg(\tilde{\gamma}_g)$ gives the exponent of $[\tilde{f}_{\tilde{x}_0}]$, because the addition between elements of the de Rham cohomology group is homomorphic to a loop product between the corresponding elements of the homotopy group.
 Moreover, since $[\tilde{f}_{\tilde{x}_0}] \in \pi_n (S^n, \tilde{x}_0)$ is isomorphic to $[f_{x_0}] \in \pi_n (S^n/K)$ for $n \ge 2$ \cite{Eilenberg;1939}, Eq. (\ref{eil-10}) is expressed in terms of the elements of the homotopy groups as
\begin{equation}
 [\gamma_g (f_{x_0})] =\gamma_g ([f_{x_0}]) = [f_{x_0}]^{\deg(\tilde{\gamma}_g)}. \label{eil-11}
\end{equation}
The above discussions are summarized in the following theorem \cite{Eilenberg;1939}:

\begin{thm}
For any $\gamma_g \in \pi_1 (S^n/K,x_0) \ \ (n \ge 2)$ that is induced by $g \in K $, the action of $\gamma_g$ on $[f_{x_0}] \in \pi_n (S^n /K,x_0)$ is given by
\begin{equation}
 \gamma_g ([f_{x_0}]) = [f_{x_0}]^{\deg(\tilde{\gamma}_g)}, \label{eil-13}
\end{equation}
where $\tilde{\gamma}_g$ is the lift of $\gamma_g$, and $\deg(\tilde{\gamma}_g) = \pm 1$.
\end{thm}\label{thm-2}
Since $\deg(\tilde{\gamma}_g) =\det (O)$, where $O \in O(n)$ is a matrix representation of $g^{\ast}$, $\deg(\tilde{\gamma}_g)$ is equal to $-1$ only when $g$ is the nontrivial element of $O(n)/SO(n) \cong \mathbb{Z}_2$. If $K=\mathbb{Z}_2 \; ( \mathbb{Z}_2 \cong O(n)/SO(n))$, the order parameter space in general satisfies the relation:
\begin{equation}
SO(n+1)/(SO(n) \rtimes \mathbb{Z}_2) \simeq S^n/\mathbb{Z}_2 \simeq \mathbb{R} \text{P}^n,
\end{equation}
where $\mathbb{R} \text{P}^n$ is an $n$-dimensional real projective space. Therefore, $g$ is equivalent to an antipodal map $g(x) = - x$, where $x \in S^n \simeq SO(n+1)/SO(n)$. The following lemma holds for the antipodal map~\cite{Greenberg:1981}:

\begin{lem}
Let $e$ and $g$ be the elements of $\mathbb{Z}_2 \cong O(n)/SO(n)$, where $g^2 = e$, and let $\gamma_e$ and $\gamma_g$ be the corresponding elements of $\pi_1(S^n/\mathbb{Z}_2,x_0)$. Then, $\deg (\gamma_e)$ and $\deg(\gamma_g)$ satisfy
 \begin{subequations}
 \begin{align}
  \deg (\tilde{\gamma}_e) &= 1,\label{eil-14-a}\\
  \deg (\tilde{\gamma}_g) &= (-1)^{n+1}. \label{eil-14}
\end{align} 
 \end{subequations}
\end{lem}\label{lem-2}
The proof of this lemma is given in, e.g., Ref. \cite{Greenberg:1981}. Lemma 1 means that the action of $O(n)/SO(n) \cong \mathbb{Z}_2$ on $S^n$ changes the orientation of the coordinates if $n$ is even. The space in which the action changes the sign of the orientation of coordinates is called a nonorientable space.

 It follows from Lemma 1 that $\deg (\widetilde{\gamma}_g) = -1$ if and only if $n$ is even and $g \in  K$ is the nontrivial element in $O(n)/SO(n) \cong \mathbb{Z}_2$. On the other hand, from Theorem 3, $\gamma_g (\alpha)$ is calculated to be $\gamma_g (\alpha) = \alpha^{\deg(\tilde{\gamma}_g)}$. Hence, we obtain the following theorem. 
\begin{thm}
For any $\gamma_g \in \pi_1 (S^n/K, x_0)$ and any $\alpha \in \pi_n (S^n /K,x_0)$ {\rm (} $n \ge 2${\rm )}, $\gamma_g (\alpha )$ is given by
\begin{equation}
 \gamma_g (\alpha) = \begin{cases} \alpha^{-1} \ \ &\text{if $n$ is even and $O(n)/SO(n) \subset K$}; \\ \alpha \ \ &\text{otherwise}. \end{cases}
\end{equation}
\end{thm}\label{thm-3}

 In other words, there is an influence of vortices for the case of $\mathcal{M} =S^n / K$ if and only if $n$ is even and $K$ includes $O(n)/SO(n) \cong \mathbb{Z}_2$.

\subsection{The case with $\mathcal{M} \simeq (U(1) \times S^n)/K'$}
\label{sec:calc-2}
We consider the case to which we can apply Eilenberg's theory. We consider the order parameter manifold given by
\begin{equation}
\mathcal{M} \simeq (U(1) \times SO(n+1)/SO(n))/K' \simeq (U(1) \times S^n)/K' \ \ (n \ge 2),
\end{equation}
where $K'$ is a discrete subgroup of $U(1) \times SO(n+1)$ whose element $g$ can be expressed as a set of elements $e^{i \theta} \in U(1) \; (0 \le \theta < 2 \pi)$ and $g_n \in SO(n+1)$. The equivalence relation imposed in $\mathcal{M} \simeq (U(1) \times S^n )/K'$ is given by $(e^{i \phi},x) \sim g(e^{i \phi},x) = (e^{i (\phi + \theta )},g_n(x))$, where $0 \le \phi < 2 \pi$ and $x \in S^n$. The homotopy groups are given by
\begin{subequations}
 \begin{align}
  \pi_1 ((U(1) \times S^n)/K', (1,x_0) ) &\cong \mathbb{Z} \times_h K', \label{S3-2-2a}\\
  \pi_k ((U(1) \times S^n)/K', (1,x_0) ) & \cong 0  \ \ (1 < k < n), \label{S3-2-2b}\\
  \pi_n ((U(1) \times S^n)/K', (1,x_0)) & \cong \mathbb{Z},  \label{S3-2-2c}
 \end{align}
\end{subequations}
Here, $(1,x_0)$ is a base point in $U(1) \times S^n$ and we have defined the $h$-product for the sake of consistency, since the right-hand side of  Eq. (\ref{S3-2-2a}) can be written as neither the direct product nor the semi-direct product of $\mathbb{Z}$ and $K'$. For any $k,l \in \mathbb{Z}$ and $g= (e^{i \theta},g_n), g'= (e^{i \theta'}, g_n') \in K'$, we define the product in $\mathbb{Z} \times_h K'$ as
\begin{equation}
 (k , g) \cdot (l,g')  = (k+l +h (g,  g' ), (e^{i (\theta + \theta' - 2 \pi h (g,g'))} , g_n g_n')). \label{S3-2-9}
\end{equation}
A map $h$ is a mapping from $K' \times K'$ to $\mathbb{Z}$ defined by
\begin{equation}
 h(g , g') = \begin{cases} 0 \ \ \text{if  } \theta + \theta' < 2 \pi; \\  1 \ \ \text{if  } \theta + \theta' \ge 2\pi.   \end{cases} \label{S3-2-10}
\end{equation}
We prove in Appendix A that $\mathbb{Z} \times_h K'$ is a group. 
By definition, $K'$ independently acts on $U(1)$ and $S^n$. Hence, for any $ g = (e^{i \theta},g_n) \in K'$, $g_n$ acts on $S^n$ without the influence of $e^{i \theta}$. Therefore, we can apply the discussion in Theorem~4 to obtain the following corollary.   
\begin{cor}
For any $\gamma_g \in \pi_1 ((U(1) \times S^n)/K', (1,x_0) )$ and any $\alpha \in \pi_n ([U(1) \times S^n] /K',(1, x_0))$ {\rm (} $n \ge 2 ${\rm )}, $\gamma_g (\alpha)$ is given by
\begin{equation}
 \gamma_g (\alpha) = \begin{cases} \alpha^{-1} \ \ &\text{if $n$ is even and $ O(n)/SO(n) \subset K'$}; \\ \alpha \ \ &\text{otherwise}. \end{cases}
\end{equation}
\end{cor}\label{cor-1}

\section{Applications to physical systems}
\label{sec:app}
In this section, we calculate the Abe homotopy group $\kappa_n$ and its equivalence classes $\kappa_n/\sim$ for the case of liquid crystals and spinor BECs. In this section, we use $n_{\rm ex}$ to denote the dimension of homotopy of topological excitations, whereas $n$ denotes the dimension of the order parameter space ($\mathcal{M} \simeq S^n/K$ or $(U(1) \times S^n)/K'$). Theorem~4 and Corollary~1 derived in Sec.~\ref{sec:calc} hold for the case of $ n_{\rm ex} =n $. 
\subsection{Liquid Crystal}
\label{sec:app-LQ}

The order parameter of a liquid crystal is described with a real symmetric tensor, $Q= \sum_{i,j=x,y,z} Q_{ij} d_i \otimes d_j$, where $\bm{d}$ is a three-dimensional unit vector, and $Q$ is a real symmetric tensor, and therefore, diagonalizable. The full symmetry of the system is $G=SO(3)$, and the free energy of the system is invariant under an arbitrary $SO(3)$ rotation of the nematic tensor $Q$. Here, an element of $M \in SO(3)$ acts on $Q$ as  
\begin{equation}
Q \mapsto M Q M^{T},
\end{equation}
where $T$ denotes transpose. 
\subsubsection{Uniaxial nematic phase}

The order parameter of the uniaxial nematic phase is a traceless symmetric tensor which is symmetric under a rotation about the $x$ axis \cite{deGenne:1993},
\begin{equation}
 \bm{Q}_{\text{UN}} = A \diag( 2/3,-1/3,-1/3),\label{app:lq-0}
\end{equation}
where $A \in \mathbb{R}$ is an amplitude of the order parameter, and diag denotes the diagonal matrix. $\bm{Q}_{\text{UN}}$ is also invariant under $\pi$ rotation about an arbitrary axis in the $yz$ plane. For example, the $\pi$ rotation about the $z$ axis is given by
\begin{align}
U =  \diag (-1,-1,1). \label{app:U} 
\end{align}
This is the case of Sec.~\ref{sec:calc-1} with $n=2$ and $K = \mathbb{Z}_2 \cong O(2)/SO(2)$. The isotropy group for the order parameter (\ref{app:lq-0}) is $H= \mathbb{Z}_2 \ltimes SO(2)  \cong O(2)$, and hence, the order parameter manifold is given by 
\begin{equation}
\mathcal{M} \simeq SO(3)/(\mathbb{Z}_2 \ltimes SO(2)  ) \simeq S^2/\mathbb{Z}_2 \simeq \mathbb{R}\text{P}^2. \label{app:lq-1}
\end{equation}
The homotopy groups with $n_{\rm ex}= 1,2$ are given by Eqs. (\ref{S3-2-2-a}) and (\ref{S3-2-2-c}), whereas for $n_{\rm ex}=3$, topological charges are given by the Hopf map $\pi_3 (S^2)$. For the case of $n_{\text{ex}} = 2$, Theorem~4 tells us that there exists an influence of vortices on $\pi_2$. We denote a nontrivial element of the fundamental group by $\gamma_U$.
Since $U \in O(2)/SO(2)$, the charge of $\pi_2$ changes to its inverse due to the influence of $\gamma_U$:
\begin{equation}
\gamma_U (\alpha) = \alpha^{-1}, \label{app:lq-4}
\end{equation}
and the equivalence relation (\ref{S3-14}) becomes 
\begin{equation}
 \alpha \sim \alpha^{-1}.\label{app:lq-5}
\end{equation}
Since $\pi_n (S^n/K,x_0) \cong \mathbb{Z} \; (n \ge 2)$ is an additive group, an arbitrary element of $\pi_n (S^n/K,x_0 )$ can be described by $\alpha^{m}$ ($m \in \mathbb{Z})$, where $\alpha \in \pi_n(S^n/K,x_0)$ characterizes  a topological excitation with a unit winding number. Therefore, the quotient group of $\pi_2 (S^2/\mathbb{Z}_2,x_0)$ under the equivalence relation of Eq.~(\ref{app:lq-5}) has only two elements, the identity and $\alpha$. As a result, the second Abe homotopy group and its equivalence classes under (\ref{app:lq-5}) are calculated to be
\begin{subequations}
\begin{align}
 &\kappa_2 (S^2/\mathbb{Z}_2, x_0) \cong \mathbb{Z}_2 \ltimes \mathbb{Z},\label{app:lq-7a} \\
 &\kappa_2 (S^2 /\mathbb{Z}_2 ,x_0)/\sim \; \cong \mathbb{Z}_2 \times \mathbb{Z}_2,\label{app:lq-7b}
 \end{align}
\end{subequations}
where the second Abe homotopy group is isomorphic to the semi-direct product of $\pi_1$ and $\pi_2$, since there is an influence of $\gamma_U$ on monopoles. When we impose the equivalence relation (\ref{S3-13}) and (\ref{S3-14}), the relation (\ref{S3-13}) is always satisfied because $\pi_1$ is a commutative group, while the subgroup $\pi_2 \subset \kappa_2$ becomes $ \mathbb{Z}_2$ by the relation (\ref{S3-14}). Moreover the semi-direct product becomes the direct product due to the relation $(\gamma_U,1_{\phi_0} ) \ast (1_{\phi_0}, \alpha) \sim (1_{\phi}, \alpha) \ast (\gamma_U, 1_{\phi_0})$. Equation (\ref{app:lq-7b}) means that there are two possibilities: a monopole exists or not. In the presence of a vortex, a pair of monopoles can be annihilated by rotating one of them around the vortex. Therefore eventually only zero or one monopole can survive. Our result shows that such pair-annihilation process occurs by creating a pair of vortex and anti-vortex even when the initial state involves no vortex.
For the case of $n_{\text{ex}} =3$, we cannot apply Theorem 4 to calculate the $\pi_1$ action on $\pi_3$, and therefore, we cannot make a definite statement as to whether there is the influence of vortices on the element of $\pi_3$.

\subsubsection{Biaxial nematic phase}
The order parameter of the biaxial nematic phase is also described by a traceless symmetric tensor~\cite{deGenne:1993},
\begin{equation}
 \bm{Q}_{\text{BN}} = \diag (A_1,A_2,A_3),\label{app:lq-15}
\end{equation}
where $A_1,A_2,A_3 \in \mathbb{R}$ are arbitrary constants that satisfy $A_1 + A_2 + A_3 = 0$. The isotropy group of the order parameter (\ref{app:lq-15}) is $H = D_2$ when $A_1 \neq A_2$, $A_2 \neq A_3$, and $A_1 \neq A_3$. The order parameter manifold is given by  
\begin{equation}
\mathcal{M} \simeq  SO(3)/D_2 \simeq SU(2)/Q_8 \simeq  S^3/Q_8, \label{app:lq-9}
\end{equation}
where we have used the fact that the universal covering space of $SO(3)$ is $SU(2)$. This lift involves a map from $D_2$ to $Q_8$, where $Q_8$ is a quaternion group $Q_8= \{ \pm \bm{1}, \pm i \sigma_x , \pm  i\sigma_y , \pm i\sigma_z \}$ with $\sigma_{\mu}$'s $(\mu=x,y,z)$ being the Pauli matrices and $\bm{1}$ is an identity. Hence, this is also the case of Sec.~\ref{sec:calc-1} but with $n=3$ and $K= Q_8$. The homotopy groups are given in Eqs. (\ref{S3-2-2-a}) -- (\ref{S3-2-2-c}). There is no nontrivial topological excitation with the dimension of homotopy $n_{\rm ex} = 2$. On the other hand, there are nontrivial topological excitations with the dimension of homotopy $n_{\rm ex}=3$. However, since $n$ is odd, Theorem 4 tells us that there is no influence of vortices. Therefore, the third Abe homotopy group is isomorphic to the direct product of $\pi_1$ and $\pi_3$: 
\begin{equation}
\begin{split}
 &\kappa_3 (S^3/Q_8, x_0 ) \cong Q_8 \times \mathbb{Z}, \\
 &\kappa_3 (S^3/Q_8, x_0 ) / \sim \ \ \cong [Q_8] \times \mathbb{Z}. \label{app:lq-13}
 \end{split}
\end{equation}
 Note that since $Q_8$ is non-Abelian, the types of vortices are classified with the conjugacy classes of $Q_8$, which is represented by \cite{Mermin;1979}
\begin{equation}
 [Q_8] = \{ \{ \bm{1}\} , \{ -\bm{1} \} ,\{ \pm i \sigma_x\} , \{ \pm i \sigma_y\} , \{ \pm i \sigma_z \} \}, \label{app:lq-14}
\end{equation}
where $\{ \cdots \}$ means that the elements in the curly brackets belong to the same conjugacy class.

\subsection{Gaseous Bose-Einstein condensates}
\label{sec:app-BEC}
A spinor BEC is a BEC of atoms with internal degrees of freedom~\cite{Ueda:arXiv}. For a spin-$F$ system, the order paramter is described by a $(2F+1)$-component spinor: 
\begin{align}
 \bm{ \Psi} = (\psi_F,\psi_{F-1},\cdots,\psi_{-F})^T,
\end{align}
where $\psi_m\, (m=F,F-1,\cdots,-F)$ describes order parameter for the magnetic sublevel $m$. The free energy of the system in the absence of the magnetic field is invariant under the $U(1)$ gauge transformation and the $SO(3)$ spin rotation, i.e., $G=U(1)_{\phi} \times SO(3)_{\bm{F}}$, where the subscripts $\phi$ and $\bm{F}$ indicate that the symmetry refers to the gauge and spin symmetry, respectively. Topological excitations in spinor BECs are discussed in~\cite{Makela:2003,Yip:2007}. We consider here spin-1 and spin-2 spinor BECs.

\subsubsection{Spin-1 BECs}
There are two phases in a spin-1 BEC: the ferromagnetic (FM) phase and polar (or antiferromagnetic) phase \cite{Ho;1998,Ohmi;1998}. The normalized order parameter for the FM phase is given by
\begin{equation}
\bm{\Psi}_{\text{ferro}} = (1,0,0)^T, \label{app:sp-2}
\end{equation}
which has the isotropy group $ H= U(1)_{\bm{F},\phi}$. The order parameter manifold is given by
\begin{equation}
\mathcal{M}\simeq (U(1)_{\phi} \times SO(3)_{\bm{F}})/U(1)_{\bm{F},\phi} \simeq SO(3)_{\bm{F},\phi} \simeq SU(2)/\mathbb{Z}_2 \simeq S^3/\mathbb{Z}_2. \label{app:sp-3}
\end{equation}
This is the case discussed in Sec. \ref{sec:calc-1} with $K = \mathbb{Z}_2 = \{ \bm{1},-\bm{1}\} \in SU(2)$ and $n=3$. In the FM phase, there is no stable topological excitation with $n_{\rm ex}=2$, while topological excitations with $n_{\rm ex}=3$ are stable and labeled with integers \cite{Khawaja;2001}. The influence of vortices on $\pi_3$ is trivial because $n$ is odd. Note that although the fundamental group is isomorphic to that of the uniaxial nematic phase in the liquid crystal, there is no influence of vortices in the present case since the dimension $n$ of the order parameter space is odd. The third Abe homotopy group is isomorphic to the direct product of $\pi_1$ and $\pi_3$: 
\begin{subequations}
\begin{align}
 &\kappa_3 (SO(3)_{\bm{F},\phi}, x_0) \cong \mathbb{Z}_2 \times \mathbb{Z}, \label{app:sp-6-a}\\
 &\kappa_3 (SO(3)_{\bm{F},\phi}, x_0)/\sim \; \cong \mathbb{Z}_2 \times \mathbb{Z}. \label{app:sp-6-b}
\end{align}
\end{subequations}

The order parameter of the polar phase is given by
\begin{equation}
\bm{\Psi}_{\text{polar}} = (0,1,0)^T, \label{app:sp-7}
\end{equation}
which has the isotropy group $H=  (\mathbb{Z}_2 )_{\bm{F}, \phi} \ltimes SO(2)_{\bm{F}} \cong O(2)_{\bm{F},\phi}$. The order parameter manifold is calculated as \cite{Fei:2001} 
\begin{equation}
\mathcal{M} \simeq (U(1) \times SO(3)_{\bm{F}})/((\mathbb{Z}_2)_{\bm{F}, \phi} \ltimes SO(2)_{\bm{F}} ) \simeq (U(1)_{\phi} \times S^2_{\bm{F}})/(\mathbb{Z}_2)_{\bm{F}, \phi}, \label{app:sp-8}
\end{equation}
and this is the case discussed in Sec. \ref{sec:calc-2} with $K' = (\mathbb{Z}_2)_{\bm{F}, \phi}$ and $n=2$. The homotopy groups with $n_{\text{ex}} =1$ and $2$ are given by Eqs. (\ref{S3-2-2a}) and (\ref{S3-2-2c}) \cite{Leonhardt;2000,Stoof:2001}, respectively, whereas the topological charge for $n_{\rm ex}=3$ is given by the Hopf map. In the polar phase, the second homotopy group has nontrivial elements, which characterizes monopoles or two-dimensional skyrmions, while the third homotopy group represents knot solitons defined by the Hopf map \cite{Kawaguchi;2008}. Since $K' = \mathbb{Z}_2 \cong O(2)_{\phi,\bm{F}}/SO(2)_{\bm{F}}$, Corollary 1 shows that there exists an influence of vortices on $\pi_2$. Hence, the second Abe homotopy group and its equivalence classes are given by
\begin{subequations}
\begin{align}
&\kappa_2 ((U(1)_{\phi} \times S^2_{\bm{F}})/ (\mathbb{Z}_2)_{\bm{F} ,\phi}, (1,x_0) )  \cong (\mathbb{Z} \times_h (\mathbb{Z}_2)_{\bm{F}, \phi}) \ltimes \mathbb{Z}, \label{app:sp-11-a}\\
 &\kappa_2 ((U(1)_{\phi} \times S^2_{\bm{F}})/ (\mathbb{Z}_2)_{\bm{F} ,\phi}, (1,x_0) ) / \sim  \; \cong  (\mathbb{Z} \times_h (\mathbb{Z}_2)_{\bm{F},\phi}) \times \mathbb{Z}_2. \label{app:sp-11-b}
\end{align}
\end{subequations}
The conjugacy class of Abe homotopy group is obtained by imposing the equivalence relations (\ref{S3-13}) and (\ref{S3-14}). The relation (\ref{S3-13}) is always satisfied because $\mathbb{Z} \times_h (\mathbb{Z}_2)_{\phi,\bm{F}}$ is an additive group, while the relation (\ref{S3-14}) reads $ \alpha \sim \alpha^{-1}$, where $\alpha \in \pi_2$. Hence, $\kappa_2 /\sim$ becomes the direct product of $\pi_1$ and $\pi_2 $. The influence of $\pi_3$, however, cannot be calculated at this stage since we cannot apply Corollary 1 to this case.

\subsubsection{Spin-2 BECs}
The order parameter of a spin-2 BEC has five complex components;
\begin{equation}
\bm{\Psi} = (\psi_2,\psi_1,\psi_0,\psi_{-1},\psi_{-2})^T. \label{app:sp-13}
\end{equation}
This system accommodates the ferromagnetic (FM), cyclic \cite{Ueda;2000,Ueda;2002,Yip;2000}, uniaxial nematic (UN), and biaxial nematic (BN) phase in the absence of the magnetic field. The phase diagram is the same as that for $d$-wave superconductors~\cite{Mermin:1974spin-2}. In the mean-field approximation, the UN and BN phases are degenerate. However, since quantum and thermal fluctuations  are known to lift the degeneracy \cite{Song;2007,Tuner;2007,Uchino;2010}, we treat these phases independently. We also consider the topological excitations in the degenerate order-parameter space. Since the influence of vortices in the FM phase is essentially the same as that in the spin-1 FM phase, we do not discuss this case.

Let us  discuss the cyclic phase. The order parameter is written as
\begin{equation}
\bm{\Psi}_{\text{cyclic}}=\frac{1}{2}(i , 0,\sqrt{2},0,i)^T, \label{app:sp-14}
\end{equation}
whose isotropy group $H$ is a tetrahedral group $T_{\bm{F}, \phi}$ \cite{Makela:2003,Semenoff;2007}. Therefore, the order parameter manifold is given by \cite{Makela:2003,Semenoff;2007}
\begin{equation}
\mathcal{M} \cong (U(1)_{\phi} \times SO(3)_{\bm{F}})/ T_{\bm{F}, \phi} \simeq (U(1)_{\phi} \times S^3_{\bm{F}})/ T^{\ast}_{\bm{F}, \phi}. \label{app:sp-15}
\end{equation}
This is the case discussed in Sec. \ref{sec:calc-2} with $K' = T^{\ast}_{\bm{F}, \phi}$ and $n=3$. Here, we lift $U(1)_{\phi} \times SO(3)_{\bm{F}}$ to $U(1) \times SU(2)$, and $T_{\bm{F},\phi}$ to $T^{\ast}_{\bm{F}, \phi}$, where $T^{\ast}_{\bm{F},\phi}$ is defined by 
\begin{multline}
  T^{\ast}_{\bm{F}, \phi} = \{ (1, \pm \bm{1}) ,  (1, \pm i \sigma_x) ,  (1, \pm i  \sigma_y)  , (1,  \pm i  \sigma_z) , \\   (e^{  \frac{2 \pi i}{3} },  \pm C_3)   , (e^{  \frac{ 2 \pi i}{3} },  \pm i \sigma_x C_3 ) , (e^{  \frac{2 \pi i}{3} }, \pm i \sigma_y C_3 ) , (e^{  \frac{2 \pi i}{3} }, \pm i \sigma_z C_3 ) , \\  (e^{  \frac{ - 2 \pi i}{3} }, \pm C_3^2 ) ,  (e^{ \frac{- 2 \pi i}{ 3}} ,\pm i \sigma_x C_3^2)  , (e^{ \frac{- 2 \pi i}{ 3}}, \pm i \sigma_y C_3^2 ),  (e^{ \frac{- 2 \pi i}{ 3}}, \pm i \sigma_z C_3^2 ) \}. \label{app:sp-16}
\end{multline}
Here, $C_3 \in SU(2)$ represents a $2 \pi/3$ rotation around a vector $ (1,1,1)$,
\begin{equation}
C_3 = \frac{1}{2}\begin{pmatrix} -1-i & -1-i \\ 1-i &-1+i  \end{pmatrix}. \label{app:sp-17}
\end{equation}
Then, homotopy groups in the cyclic phase are given in Eqs. (\ref{S3-2-2a}) -- (\ref{S3-2-2c}). The influence of vortices on $\pi_3$ does not exist because $n$ is odd. Thus, the third Abe homotopy group becomes the direct product of $\pi_1$ and $\pi_3$:
\begin{subequations}
\begin{align}
 &\kappa_3 ((U(1)_{\phi} \times SO(3)_{\bm{F}})/T_{\bm{F},\phi},(1,x_0) ) \cong ( \mathbb{Z} \times_h {T}^{\ast}_{\bm{F},\phi}) \times \mathbb{Z}, \label{app:sp-20-a}\\
 &\kappa_3 ((U(1)_{\phi} \times SO(3)_{\bm{F}})/T_{\bm{F},\phi},(1,x_0) )/\sim \; \cong [ \mathbb{Z} \times_h {T}^{\ast}_{\bm{F},\phi}] \times \mathbb{Z}, \label{app:sp-20-b}
\end{align}
\end{subequations}
where $[\mathbb{Z} \times_h T^{\ast}_{\bm{F}, \phi}]$ is the conjugacy class of $\mathbb{Z} \times_h T^{\ast}_{\bm{F}, \phi}$ whose elements are listed in \ref{app-b}. Equations (\ref{app:sp-20-a}) and (\ref{app:sp-20-b}) show that vortices and topological excitations of $\pi_3$ are topologically independent.
 
The order parameter in the UN phase is given by
\begin{equation}
 \bm{\Psi}_{\text{UN}} = (0,0,1,0,0)^T.\label{app:sp-22}
\end{equation}
The isotropy group of the UN phase is $(\mathbb{Z}_2)_{\bm{F}} \ltimes SO(2)_{\bm{F}} \cong O(2)_{\bm{F}}$. Therefore, the order parameter space is given by \cite{Song;2007}
\begin{equation}
\mathcal{M} \simeq U(1)_{\phi} \times S^2_{\bm{F}}/(\mathbb{Z}_2)_{\bm{F}}, \label{app:sp-23}
\end{equation}
and the results in Sec. \ref{sec:calc-2} are applicable, where $K' = (\mathbb{Z}_2)_{\bm{F}}$ and $n=2$. The homotopy group with $n_{\rm ex}=1,2,$ and $3$ are given by Eq. (\ref{S3-2-2a}), (\ref{S3-2-2c}), and the Hopf map $\pi_3 (S^2)$, respectively. In this case, since $\mathbb{Z}_2$ does not include elements of $U(1)$, the fundamental group becomes the direct product of $\mathbb{Z}$ and $\mathbb{Z}_2$ rather than the $h$-product. Therefore, the calculation of the influence of vortices can be carried out in a manner similar to the case of $S^2/\mathbb{Z}_2$. Since $K' = \mathbb{Z}_2 \cong O(2)_{\bm{F}}/SO(2)_{\bm{F}}$, the second Abe homotopy group is a nontrivial semi-direct product of $\pi_1$ and $\pi_2$:
\begin{subequations}
\begin{align}
& \kappa_2 (U(1)_{\phi} \times S^2_{\bm{F}}/(\mathbb{Z}_2)_{\bm{F}},x_0) \cong \mathbb{Z} \times \mathbb{Z}_2 \ltimes \mathbb{Z}, \label{app:sp-26-a}\\
& \kappa_2 (U(1)_{\phi} \times S^2_{\bm{F}}/(\mathbb{Z}_2)_{\bm{F}},x_0) /\sim \; \cong \mathbb{Z} \times \mathbb{Z}_2 \times \mathbb{Z}_2, \label{app:sp-26-b}
 \end{align}
\end{subequations}
where the semi-direct product of $\pi_1$ and $\pi_2$ reduces to $\mathbb{Z}_2 \times \mathbb{Z}_2$ by the equivalence relation. 
The third Abe homotopy group, however, cannot be calculated in this manner because Theorem 4 is not applicable.

Next, the order parameter of the BN phase is given by
\begin{equation}
 \bm{\Psi}_{BN} = \frac{1}{\sqrt{2}} (1,0,0,0,1)^T, \label{app:sp-28}
\end{equation}
whose isotropy group is $H= (D_4)_{\bm{F}, \phi}$, where $(D_4)_{\bm{F}, \phi}$ is the fourth dihedral group, and the order parameter manifold is given by \cite{Song;2007}
\begin{equation}
\mathcal{M} \simeq (U(1)_{\phi} \times SO(3)_{\bm{F}})/(D_4)_{\bm{F}, \phi} \simeq (U(1)_{\phi} \times S^3_{\bm{F}})/(D_4^{\ast})_{\bm{F}, \phi}. \label{app:sp-29}
\end{equation}
When we lift $U(1) \times SO(3)$ to $U(1) \times SU(2)$, $(D_4)_{\bm{F}, \phi}$ is lifted to $({D_4}^{\ast})_{\bm{F}, \phi}$:
\begin{multline}
 ({D_4}^{\ast})_{\bm{F}, \phi} = \{ (1, \pm \bm{1}) , (1, \pm i \sigma_x) , (1, \pm  i \sigma_y) , (1, \pm  i \sigma_z) , \\ (e^{ i \pi },\pm C_4)  ,  ( e^{ i \pi },\pm C_4^3)   , ( e^{ i \pi },\pm i \sigma_x C_4)  ,(  e^{ i  \pi }, \pm i \sigma_x C_4^3) \},
\end{multline}
where $C_4 \in SU(2)$ represents a $\pi/2$ rotation about the $z$ axis, which is given by
\begin{equation}
 C_4 = \frac{1}{\sqrt{2}} \begin{pmatrix} 1 + i & 0 \\ 0 & 1- i\end{pmatrix}. \label{app:sp-32}
\end{equation}
This is the case discussed in Sec. \ref{sec:calc-2} with $K' = ({D_4}^{\ast})_{\bm{F}, \phi}$ and $n=3$. Therefore, homotopy groups in the BN phase are given by Eqs. (\ref{S3-2-2a}) -- (\ref{S3-2-2c}).
In the BN phase, there is no topological excitation with dimension of homotopy $n_{\rm ex} = 2$, whereas there are three-dimensional skyrmions labeled by elements of $\pi_3$. The influence of vortices on $\pi_3$, however, is trivial since $n$ is odd. Hence, the third Abe homotopy group is given by
\begin{subequations}
\begin{align}
& \kappa_3 ( (U(1)_{\phi} \times SO(3)_{\bm{F}})/(D_4)_{\bm{F}, \phi},(1,x_0)) \cong (\mathbb{Z} \times_h (D_4^{\ast})_{\bm{F},\phi}) \times  \mathbb{Z}, \label{app:sp-34-a}\\
& \kappa_3 ( (U(1)_{\phi} \times SO(3)_{\bm{F}})/(D_4)_{\bm{F}, \phi},(1, x_0)) /\sim \; \cong [\mathbb{Z} \times_h (D_4^{\ast})_{\bm{F},\phi}] \times  \mathbb{Z} , \label{app:sp-34-b}
\end{align}
\end{subequations}
where the third Abe homotopy group becomes the direct product of $\pi_1$ and $\pi_3$. Here, $[\mathbb{Z} \times_h (D_4^{\ast})_{\bm{F},\phi}]$ is the conjugacy class of $\mathbb{Z} \times_h (D_4^{\ast})_{\bm{F},\phi}$, which is shown in \ref{app-b}. According to (\ref{app:sp-34-a}), vortices and three-dimensional skyrmions are independent of each other in the BN phase. 

Finally, let us consider the case in the absence of quantum fluctuations. In this case, the UN and BN phases are degenerate. The order parameter that describes both phases is given by \cite{Tuner;2007,Uchino;2010} 
\begin{equation}
\bm{\Psi} = (\sin \eta / \sqrt{2}, 0 , \cos \eta , 0 , \sin \eta /\sqrt{2})^T, \label{app:sp-36}
\end{equation}
where $\eta$ varies from $0$ to $\pi/3$. When $\eta = \pi/3$, $\mathcal{M}$ is the order parameter space of the UN phase. When $\eta =\pi/6$, $\mathcal{M}$ is the order parameter of the BN phase. Otherwise, they have the isotropy group $H = (D_2)_{\bm{F}}$. The degree of freedom of $\eta$ originates from the fact that the nematic phase has the accidental symmetry bigger than other phases, resulting in the emergence of the  quasi-Nambu-Goldstone mode \cite{Uchino;2010} \footnote{  The accidental symmetry, resulting in the emergence of
quasi-Nambu-Goldstone mode, also occurs in superfluid $^3$He-A~\cite{Volovic:1982,Novikov:1982,Brauner:2010} and liquid crystals. In these systems, the homotopy classification has been applied for the investigation of core structures of vortices~\cite{Mermin:1978qng,Lyuksyutov:1978}.  The stability of vortices in theories with quasi-Nambu-Goldstone modes has been also studied in the context of supersymmetric gauge theories~\cite{Penin:1996,Achucarro:2002}.}. The full symmetry of the nematic phase is $G=U(1)_{\phi} \times SO(5)_{A_{20}}$, where subscript $A_{20}$ represents the symmetry preserving a singlet-pair amplitude.

The isotropy group of the nematic phase is  $H= (\mathbb{Z}_2)_{A_{20},\phi} \ltimes SO(4)_{A_{20}} \cong O(4)_{A_{20}, \phi}$. Thus, the order parameter manifold is given by \cite{Uchino;2010}
\begin{equation}
\mathcal{M} \simeq (U(1)_{\phi} \times SO(5)_{A_{20}})/ ((\mathbb{Z}_2)_{A_{20},\phi} \ltimes SO(4)_{A_{20}}) \simeq (U(1)_{\phi} \times S^4_{A_{20}})/ (\mathbb{Z}_2)_{A_{20},\phi}. \label{app:sp-37}
\end{equation}
This corresponds to the case discussed in Sec.~\ref{sec:calc-2} with $K' = (\mathbb{Z}_2)_{A_{20},\phi}$ and $n=4$. The conventional homotopy groups in the nematic phase are given in Eqs.~(\ref{S3-2-2a}) -- (\ref{S3-2-2c}). In the nematic phase, $\pi_2$ and $\pi_3$ are trivial and thus a point defect and a three-dimensional skyrmion are unstable. However, there is a nontrivial element of $\pi_4$, which is given by   
\begin{equation}
 \pi_4 ( (U(1)_{\phi} \times S^4_{A_{20}})/ (\mathbb{Z}_2)_{A_{20},\phi}, (1,x_0)) \cong \mathbb{Z}, \label{app:sp-39}
\end{equation}
where $\pi_4$ is interpreted as a describing four-dimensional texture called an instanton. The instanton is point-like in the four-dimensional Euclidean space including time. Since $n$ is even and $K' = \mathbb{Z}_2 \cong O(4)_{A_{20}, \phi}/SO(4)_{A_{20}}$, from Corollary 1, the influence of vortices exists.   
 The influence of votrices restricts the instanton charge to $\mathbb{Z}_2$ because the equivalence relation (\ref{S3-14}) gives $\beta \sim \beta^{-1} \; (\beta \in \pi_4) $. The fourth Abe homotopy group and its conjugacy class are given by
\begin{subequations}
\begin{align}
 &\kappa_4 ((U(1)_{\phi} \times S^4_{A_{20}})/ (\mathbb{Z}_2)_{A_{20}, \phi},(1,x_0))  \cong (\mathbb{Z} \times_h (\mathbb{Z}_2)_{A_{20}, \phi} )\ltimes \mathbb{Z}, \label{app:sp-41-a}\\
 &\kappa_4 ((U(1)_{\phi} \times S^4_{A_{20}})/ (\mathbb{Z}_2)_{A_{20}, \phi},(1,x_0)) / \sim \; \cong (\mathbb{Z} \times_h (\mathbb{Z}_2)_{A_{20}, \phi}) \times \mathbb{Z}_2. \label{app:sp-41-b}
 \end{align}
\end{subequations}
Therefore, the instanton charge by $\pi_4$ cannot be defined uniquely due to the influence of vortices. The real charge should be defined by the conjugacy classes of the fourth Abe homotopy group. 

The topological charge under the influence of vortices is summarized in Table~\ref{inf-table}, in which topological excitations in superfluid helium 3 are also classified. The order parameter space for superfluid helium 3 is given in Ref.~\cite{Volovik;1977inf}. The homotopy groups and the conjugacy classes of Abe homotopy groups are listed in Table.~\ref{Abe-table}.
\begin{table}[btph]
\centering
 \caption{Topological charge for the dimension of homotopy $n_{\rm ex}=2$ and $3$ in the presence of vortices. Here, UN, BN, and FM stand for uniaxial nematic, biaxial nematic, and ferromagnetic phases, respectively. In superfluid $^3$He, we classify $^3$He-B, $^3$He-A, and  $^3$He-A$_1$ phases.  When there exists the influence of vortices, we show how the topological charges due to the influence of vortices. For example, for the case of the UN phase in a liquid crystal, $\mathbb{Z} \to \mathbb{Z}_2$ in the column of $n_{\text{ex}} =2$ indicates that although $\pi_2 \cong \mathbb{Z}$, the topological invariant which describes a monopole is reduced to $\mathbb{Z}_2$ due to the influence of vortices. When it is $\mathbb{Z}$, there is no influence of vortices. Here, $0$ indicates that there is no nontrivial topological excitation.  \vspace*{2mm}}\label{inf-table}
 \begin{tabular}{cccc} \hline \hline
 system & phase & $n_{\rm ex}=2$ & $n_{\rm ex}=3$  \\ \hline
  liquid crystal & UN & $\mathbb{Z} \to \mathbb{Z}_2$ & ?  \\
                       & BN & 0 & $\mathbb{Z}$  \\
  gaseous BEC & spin-1 FM & 0 & $\mathbb{Z}$  \\
                       & spin-1 polar & $\mathbb{Z} \to \mathbb{Z}_2$ & ?  \\
                       & spin-2 FM & 0 & $\mathbb{Z}$  \\
                       & spin-2 cyclic & 0 & $\mathbb{Z}$  \\
                       & spin-2 UN & $\mathbb{Z} \to \mathbb{Z}_2$ & ?  \\
                       & spin-2 BN & 0 & $\mathbb{Z}$  \\
 $^3$He-B        & dipole-free & 0 & $\mathbb{Z}$  \\
                      & dipole-locked & $\mathbb{Z}$ & $\mathbb{Z}$  \\
 $^3$He-A       & dipole-free & $\mathbb{Z} \to \mathbb{Z}_2$ & ?  \\
                     & dipole-locked & 0 & $\mathbb{Z}$  \\
 $^3$He-A$_1$  & dipole-free & $\mathbb{Z}$ & $\mathbb{Z}$  \\                    
                                           & dipole-locked & 0 & 0 \\  \hline \hline
 \end{tabular}
\end{table}

%\begin{rotate}{90}
\begin{sidewaystable}[hp]
\caption{
Classification of the Abe homotopy group for liquid crystals, gaseous Bose-Einstein condensates (BECs), and superfluid $^3$He-A, $^3$He-A$_1$, and $^3$He-B phases \cite{Mineev;1998,Volovik;2003}, where $[\cdots]$ denotes the conjugacy class, $Q_8$ is the quaternion group, and $D_n$ and $T$ are the $n$th dihedral group and tetrahedral group, respectively. The subscripts $\phi$, $\bm{F}$, and $A_{20}$ refer to the gauge symmetry, the hyperfine-spin symmetry, and the symmetry preserving the spin-singlet amplitude, respectively. The subscript of $\bm{S}$ and $\bm{L}$ describe the spin and orbital symmetries. We denote a vortex as $\mathbb{Z} \times_h (K)_{\bm{F}, \phi}$ in spinor BECs, where group $K$ is constructed based on the composite symmetry between the gauge symmetry $\phi$ and the hyperfine- spin symmetry $\bm{F}$. For any $n,m \in \mathbb{Z}, g,g' \in K$, $(n,g) \in \mathbb{Z} \times_h (K)_{\bm{F}, \phi}$ satisfies that $(n,g)\cdot  (m,g') = (n+m+h(g \cdot g') ,g \cdot g')$, where map $h: K \times K\to \mathbb{Z}$ is defined such that $h(g , g')= 0$ when $\theta + \theta' < 2 \pi$ and $h(g, g')= 1$ when $\theta + \theta' \ge 2 \pi$.  $K^{\ast}$ is defined as $f(K) :=K^{\ast}$ by the map $f: U(1) \times SO(3) \to U(1) \times SU(2)$.  In the nematic phase of a spin-2 BEC, there is a nontrivial influence on $\pi_4$, which is classified by the fourth Abe homotopy group: $\kappa_4([U(1)_{\phi} \times S^4_{A_{20}}]/(\mathbb{Z}_2)_{A_{20} ,\phi},(1,x_0)) = \mathbb{Z} \times_h (\mathbb{Z}_2)_{\bm{F} ,\phi} \times \mathbb{Z}_2$.      
\vspace*{2mm}  }\label{Abe-table}
 \centering
  \begin{tabular}{cccccccc} \hline \hline
 system  &phase                             & $\mathcal{M}$                                        & $\pi_1$           & $\pi_2$        &       $\pi_3$         & $\kappa_2/\sim$                                  & $\kappa_3/\sim$                             \\    \hline  %& $\kappa_4/\sim$\\ \hline
 
 % \multicolumn{5}{l}{Liquid crystal} \\ %\hline 
  
 liquid   crystal & UN                    & $\mathbb{R}$P$^2$ \cite{Mermin;1979,Michel;1980}                                &$\mathbb{Z}_2$     & $\mathbb{Z}$ & $\mathbb{Z}$   &$\mathbb{Z}_2 \times \mathbb{Z}_2$       & ?   \\ %& $\mathbb{Z}_2 \times \mathbb{Z}_2$   \\ % \hline 
 & BN             & $SO(3)/D_2$ \cite{Mermin;1979,Michel;1980}                      & $Q_8 $    & $ 0 $&  $\mathbb{Z}$      & $[Q_8] \times 0 $                   & $[Q_8] \times \mathbb{Z}$            \\ %& $[\mathbb{Q}] \times \mathbb{Z}_2$  \\ \hline \hline

 % \multicolumn{5}{l}{ BECs} \\ %\hline

gaseous  BEC& scalar                     & U(1)                                                       &$\mathbb{Z}$     & $0$ &   $0$    & $\mathbb{Z}\times 0$                    &$\mathbb{Z} \times 0$                 \\     %&$\mathbb{Z}\times \{1\}$          \\% \hline
 & spin-1 FM            & $SO(3)_{\bm{F}, \phi }$ \cite{Ho;1998}                                                  &$\mathbb{Z}_2$   &$ 0 $&  $\mathbb{Z}$     &$\mathbb{Z}_2 \times 0$                   & $\mathbb{Z}_2 \times \mathbb{Z}$      \\ %  & $\mathbb{Z}_2 \times \mathbb{Z}_2$     \\ %\hline
 & spin-1 polar               & $(U(1)_{\phi} \times S^2_{\bm{F}})/(\mathbb{Z}_2)_{\bm{F} , \phi}$ \cite{Fei:2001}     &$\mathbb{Z} \times_h (\mathbb{Z}_2)_{\bm{F} ,\phi}$     & $\mathbb{Z}$& $\mathbb{Z}$     & $\mathbb{Z} \times_h (\mathbb{Z}_2)_{\bm{F} ,\phi} \times \mathbb{Z}_2$         &   ? \\%$(\mathbb{Z})_{\phi} \times_h (\mathbb{Z}_2)_{\bm{F} ,\phi} \times \mathbb{Z}$            \\     & $\mathbb{Z} /2  \times \mathbb{Z}_2$    \\ %\hline
 & spin-2 FM & $SO(3)_{\bm{F}, \phi}/ (\mathbb{Z}_2)_{\bm{F}, \phi}$ \cite{Makela:2003}& $\mathbb{Z}_4$  &  $ 0 $  & $\mathbb{Z}$ & $\mathbb{Z}_4 \times 0 $ & $\mathbb{Z}_4 \times \mathbb{Z} $ \\
 & spin-2 nematic          & $(U(1)_{\phi} \times S^4_{A_{20}})/(\mathbb{Z}_2)_{A_{20} ,\phi}$ \cite{Uchino;2010}     &$\mathbb{Z} \times_h (\mathbb{Z}_2)_{\bm{F},\phi}$    &$0$& $0$     & $\mathbb{Z} \times_h (\mathbb{Z}_2)_{\bm{F},\phi} \times 0$                     & $\mathbb{Z} \times_h (\mathbb{Z}_2)_{\bm{F},\phi} \times 0$               \\   %   & $\mathbb{Z} /2  \times \mathbb{Z}_2$ \\% \hline
 %& spin-2 nematic (with LHY correction)  & $U(1)_{\phi} \times SO(3)_{\bm{F}}/(D_2)_{\bm{F}}$ & $\mathbb{Z} \times \mathbb{Q}_8$   & $\{ 1\}$&  $\mathbb{Z}$ &  $\mathbb{Z} \times [\mathbb{Q}_8] \times \{ 1 \} $&  $\mathbb{Z} \times [\mathbb{Q}_8] \times \mathbb{Z}$ \\
 
 &  spin-2 UN  & $U(1)_{\phi} \times S^2_{\bm{F}}/(\mathbb{Z}_2)_{\bm{F}}$ \cite{Song;2007} &  $\mathbb{Z} \times \mathbb{Z}_2$ & $\mathbb{Z}$ & $\mathbb{Z}$ & $\mathbb{Z} \times \mathbb{Z}_2 \times \mathbb{Z}_2 $ & ? \\
& spin-2 BN & $(U(1)_{\phi} \times SO(3)_{\bm{F}})/(D_4)_{\bm{F},\phi} $ \cite{Song;2007} & $\mathbb{Z} \times_h (D_4^{\ast})_{\bm{F},\phi}$ & $0$ & $\mathbb{Z}$ & $ [\mathbb{Z} \times_h (D_4^{\ast})_{\bm{F},\phi}] \times 0 $ &$[\mathbb{Z} \times_h (D_4^{\ast})_{\bm{F},\phi}] \times \mathbb{Z}$ \\

 & spin-2 cyclic            & $(U(1)_{\phi} \times SO(3)_{\bm{F}})/T_{\bm{F},\phi}$  \cite{Semenoff;2007,Makela:2003}   & $\mathbb{Z} \times_h T^{\ast}_{\bm{F},\phi}$  &  $0$& $\mathbb{Z}$ & $[\mathbb{Z} \times_h (T^{\ast})_{\bm{F},\phi}] \times 0$          & $[\mathbb{Z} \times_h (T^{\ast})_{\bm{F},\phi}] \times \mathbb{Z}$  \\ % & $[\mathbb{T}^{\ast}] \times \mathbb{Z}_2$  \\ %\hline 

  %\multicolumn{5}{l}{ $^3$He-A} \\ %\hline
 
  $^3$He-B & dipole-free        & $U(1)_{\phi} \times SO(3)_{\bm{L} + \bm{S} } $\cite{Volovik;1977inf}   & $\mathbb{Z} \times \mathbb{Z}_2 $ & $0$ & $\mathbb{Z}$  & $\mathbb{Z} \times \mathbb{Z}_2 \times 0 $ & $\mathbb{Z} \times \mathbb{Z}_2 \times \mathbb{Z}$ \\
               & dipole-locked        & $U(1)_{\phi} \times S^2_{\bm{L} + \bm{S}}$ \cite{Volovik;1977inf}         & $\mathbb{Z}$ & $\mathbb{Z}$ & $\mathbb{Z} $  &$\mathbb{Z} \times \mathbb{Z}$ & $\mathbb{Z} \times \mathbb{Z}$ \\

$^3$He-A &  dipole-free              & $(S^2_{\bm{S}} \times SO(3)_{\bm{L}})/(\mathbb{Z}_2)_{\bm{L},\bm{S}}$ \cite{Volovik;1977inf}   &$\mathbb{Z}_4$        & $\mathbb{Z}$ & $\mathbb{Z} \times \mathbb{Z}$ & $\mathbb{Z}_4 \times \mathbb{Z}_2$     & ? \\  %& $\mathbb{Z}_4 \times \mathbb{Z}_2 \times \mathbb{Z}_2$   \\ %\hline
 &  dipole-locked           & $SO(3)_{\bm{L},\bm{S}}$ \cite{Volovik;1977inf}               &$\mathbb{Z}_2$     &  $ 0$ &  $\mathbb{Z}$   &$\mathbb{Z}_2 \times 0$                  & $\mathbb{Z}_2 \times \mathbb{Z} $       \\ % & $\mathbb{Z}_2 \times \mathbb{Z}_2$      \\ %\hline
  $^3$He-A$_1$  & dipole-free &  $  U(1)_{\phi, L_z ,S_z} \times S^2_{\bm{L}}$ \cite{Volovik;1977inf} & $\mathbb{Z}$ &$\mathbb{Z} $ & $\mathbb{Z} $&$\mathbb{Z} \times \mathbb{Z} $ & $ \mathbb{Z} \times \mathbb{Z} $ \\  
   & dipole-locked  & $U(1)_{\phi, L_z,S_z}$ \cite{Volovik;1977inf}  & $\mathbb{Z}  $ &  $ 0$& $ 0$ & $\mathbb{Z}  \times 0$ & $\mathbb{Z} \times 0 $ \\
 \hline \hline 
                  
  \end{tabular} 
%\end{figure}
%\end{table}
\end{sidewaystable}
%\end{rotate}

\section{Summary and Concluding Remarks}
\label{sec:sum}
 In this paper, we provide the classification of topological excitations with the dimension of homotopy $n \ge 2$ based on the Abe homotopy group. When there exist vortices, $\pi_n$ does not give a consistent topological charge. The Abe homotopy group is applicable to classify topological excitations in the presence of vortices and its conjugacy classes give a physically consistent charge. We have shown the following results: 
\begin{itemize}
\item The conjugacy classes of the Abe homotopy group is equivalent to the type of topological excitations that coexist with vortices. 
\item  The vortex and anti-vortex pair creation, even if it is a virtual process, gives the influence on topological excitations, and, therefore, the influence of vortices always exists.
\item The relationship between a vortex and a topological excitation with the dimension of homotopy $n \ge 2$ is determined solely by the noncommutativity of elements between $\pi_1 (\mathcal{M},\phi_0)$ and $\pi_n (\mathcal{M},\phi_0)$.    
\item  If the order parameter manifold is $ SO(n+1)/(K \ltimes SO(n)) \; (K \subset SO(n+1) )$ or $(U(1) \times SO(n+1)/SO(n))/K' \; (K' \subset U(1) \times SO(n+1))$, there exists an influence of vortice on $\pi_n$ if and only if $n$ is even, and either $K$ or $K'$ includes $O(n)/SO(n) \cong \mathbb{Z}_2$.
\end{itemize}

We summarize the main results of each section. In Sec \ref{sec:review}, we showed that the topological charge under the influence of vortices is determined by the conjugacy classes of the Abe homotopy group. The influence of vortices corresponds to the nontrivial semi-direct product in the Abe homotopy group. In order to define the physically consistent charges, we take the conjugacy class not only of elements between $\pi_1(\mathcal{M} ,\phi_0)$ but also of elements between $\pi_1(\mathcal{M} ,\phi_0)$ and $\pi_n(\mathcal{M} ,\phi_0)$. The equivalence relations lead to $\gamma (\alpha) \sim \alpha$ for $\gamma \in \pi_1 (\mathcal{M},\phi)$ and $\alpha \in \pi_n (\mathcal{M},\phi)$. Here, $\gamma (\alpha)$ describes the $\pi_1(\mathcal{M} ,\phi_0)$ action on $\pi_n(\mathcal{M} ,\phi_0)$.

In Sec. \ref{sec:calc}, we developed the method to calculate $\gamma (\alpha)$ by using the differential form of Eilenberg's theory. As a result, we proved that $\pi_n (S^n / K,x_0)$ or $\pi_n ((U(1)\times S^n)/K',(1,x_0))$ has the nontrivial influence of vortices only if $n $ is even and either $K$ or $K'$ includes $O(n)/SO(n) \cong \mathbb{Z}_2$.  

In Sec. \ref{sec:app}, we calculated the Abe homotopy group for liquid crystals, gaseous spinor BECs, and the $^3$He-A and B phases \cite{Volovik;2003}. We classfied the topological excitation with the dimension of homotopy $n \ge 2$, and showed that the influence exists for the following systems: $\pi_2$ of the nematic phase of the liquid crystal \cite{Mermin;1978inf}, $\pi_2$ of the $^3$He-A dipole-free state \cite{Volovik;1977inf}, $\pi_2 $ of the spin-1 polar phase and the spin-2 UN phase, and $\pi_4$ of the spin-2 nematic phase in gaseous BECs. Here, $\pi_4$ describes the instanton, whose charge is reduced from $\mathbb{Z}$ to $\mathbb{Z}_2$ due to the influence of vortices.

An important finding obtained by using the Abe homotopy group is that the vortex and anti-vortex pair creation process also gives the influence on topological excitations. Because such a pair can be created in a virtual process, the influence always exists.This consequence might be relevant to the monopole problem in GUTs. If an Alice cosmic string exists in the Universe \cite{Preskill;1979zi}, the monopole charge is reduced to $\mathbb{Z}_2$. Thus, the Abe homotopy group may provide a new insight into the monopole problem.    

This work was supported by Grants-in-Aid for Scientific Research (Kakenhi 22340114, 22103005, 22740265), a Grant-in-Aid for scientific Research on Innovative Areas "Topological Quantum Phenomena" (No. 22103005, 23103515), a Global COE Program ``the Physical Science Frontier", and the Photon Frontier Netowork Program of MEXT of Japan. S.K. acknowledges support from JSPS (Grant No.228338).

\appendix
\section{The group property of $\mathbb{Z} \times_h K'$}
\label{app-a}
We show that the product in $\mathbb{Z} \times_h K'$ defined in Eqs. (\ref{S3-2-9}) and (\ref{S3-2-10}) satisfies the associative property and has the inverse. Here, $K'$ is a subgroup of $U(1) \times SO(n+1)$. Let $ (k , g)$, $(l,g')$, and $(m, g'')$ be elements of $\mathbb{Z} \times_h K'$ where $k,l,m \in \mathbb{Z}$ and $g=(e^{i \theta},g_n), g'=(e^{i \theta'}, g_n'),g''=(e^{i \theta''},g_n'') \in K'$ with $0 \le \theta , \theta' , \theta'' < 2 \pi$ and $g_n , g_n' , g_n'' \in SO(n+1)$. First, we verify the associative law. Following the definition of the $h$-product given in Eq. (\ref{S3-2-9}), the product of three elements of $\mathbb{Z} \times_h K'$ is calculated as follows:
\begin{subequations}
\begin{align}
 &(k,g) \cdot ((l, g') \cdot (m,g'')) \notag \\
&= (k+l+m+ h(g',g'') + h(g,g'g''), (e^{i(\theta + \theta' + \theta'' - 2 \pi h(g',g'') - 2 \pi h(g,g'g''))}, g_n g_n'g_n'')), \label{appA1-a}\\
 & ((k,g) \cdot (l, g')) \cdot (m,g'') \notag \\
&= (k+l+m+ h(g,g') + h(gg',g''), (e^{i(\theta + \theta' + \theta'' - 2 \pi h(g,g') - 2 \pi h(gg',g''))}, g_n g_n'g_n'')).\label{appA1-b}
  \end{align}
\end{subequations}
We compare these products for all possible combinations of $\theta, \theta'$ and $\theta''$.
\begin{subequations}
\begin{align}
  (a) \ \ \theta + \theta' \ge 2 \pi, \; &\theta' + \theta'' \ge 2 \pi,  \theta + \theta'' \ge 2 \pi, \label{appA2-a}\\
  (b) \ \ \theta + \theta' \ge 2 \pi, \; &\theta' + \theta'' < 2 \pi,     \theta + \theta'' \ge 2 \pi, \label{appA2-b}\\
  (c) \ \ \theta + \theta' < 2 \pi,    \; &\theta' + \theta'' \ge 2 \pi,  \theta + \theta'' \ge 2 \pi, \label{appA2-c}\\
  (d) \ \ \theta + \theta' \ge 2 \pi, \; &\theta' + \theta'' \ge 2 \pi,  \theta + \theta'' < 2 \pi,  \label{appA2-d}\\
  (e) \ \ \theta + \theta' \ge 2 \pi, \; &\theta' + \theta'' < 2 \pi,     \theta + \theta'' < 2 \pi, \label{appA2-e}\\
  (f) \ \ \theta + \theta' < 2 \pi,    \; &\theta' + \theta'' \ge 2 \pi,  \theta + \theta'' < 2 \pi, \label{appA2-f}\\
  (g) \ \ \theta + \theta' < 2 \pi,    \; &\theta' + \theta'' < 2 \pi,     \theta + \theta'' \ge 2 \pi, \label{appA2-g}\\
  (h) \ \ \theta + \theta' < 2 \pi,    \; &\theta' + \theta'' < 2 \pi,    \theta + \theta'' < 2 \pi, \; \theta + \theta' + \theta'' \ge 2 \pi, \label{appA2-h}\\
  (i) \ \ \theta + \theta' < 2 \pi,    \; &\theta' + \theta'' < 2 \pi,    \theta + \theta'' < 2 \pi, \; \theta + \theta' + \theta'' < 2 \pi. \label{appA2-i}
 \end{align}
 \end{subequations}
 In the following, we calculate $h(g,g')$, $h(g',g'')$, $h(gg',g'')$, and $h(g,g'g'')$ to show that the relation
\begin{align}
 h(g,g') + h(gg',g'') = h(g',g'') + h(g,g'g''), \label{appA3}
\end{align}  
is satisfied for all cases.
\begin{itemize}
 \item[(a)] From $\theta + \theta' \ge 2 \pi$ and $\theta' + \theta'' \ge 2 \pi$, we obtain $h(g,g')=1$ and $h(g',g'')=1$. Since $gg' = (e^{i (\theta + \theta')}, g_n g_n')$ and $2 \pi \le \theta + \theta' < 4 \pi$, we should subtract $2 \pi$ from $\theta + \theta'$. Thus, $h(gg', g)$ is given by
\begin{equation}
 h(gg' , g'') = \begin{cases} 0 \ \ \text{if  } \theta + \theta'+\theta'' - 2 \pi < 2 \pi ; \\  1 \ \ \text{if  } \theta + \theta' + \theta'' - 2 \pi \ge 2\pi.   \end{cases}
\end{equation} 
Similarly, $h(g,g'g'')$ is given by
\begin{equation}
 h(g , g' g'') = \begin{cases} 0 \ \ \text{if  } \theta + \theta'+\theta'' - 2 \pi < 2 \pi ; \\  1 \ \ \text{if  } \theta + \theta' + \theta'' - 2 \pi \ge 2\pi.   \end{cases} 
\end{equation} 
Therefore, $h(gg' , g'')$ is equivalent to $h(g , g' g'')$, and we obtain Eq. (\ref{appA3}).

\item[(b),(e)] From  $\theta + \theta' \ge 2 \pi$ and $\theta' + \theta'' < 2 \pi$, we obtain $h(g,g')=1$ and $h(g',g'')=0$. By the inequality $\theta' + \theta'' < 2 \pi$, we obtain $\theta + \theta' + \theta'' - 2 \pi < \theta < 2 \pi$. Hence, we obtain $h (gg',g'') = 0$. On the other hand, since we obtain $\theta + \theta' + \theta'' \ge 2 \pi$ and $\theta' + \theta'' < 2 \pi$ , we get $h(g, g'g'') = 1$.

\item[(c),(f)] From  $\theta + \theta' < 2 \pi$ and $\theta' + \theta'' \ge 2 \pi$, we obtain $h(g,g')=0$ and $h(g',g'')=1$. Similarly to (b), by $\theta + \theta' < 2 \pi$, we obtain $\theta + \theta' + \theta'' - 2 \pi < \theta'' < 2 \pi$. We can get $h (g,g'g'') = 0$ because of $2 \pi \le \theta' + \theta'' < 4 \pi$. On the other hand, since we obtain $\theta + \theta' + \theta'' \ge 2 \pi$ and $\theta + \theta' < 2 \pi$ , we get $h(gg', g'') = 1$.

\item[(d)] From  $\theta + \theta' \ge 2 \pi$ and $\theta' + \theta'' \ge 2 \pi$, we obtain $h(g,g')=1$ and $h(g',g'')=1$. Since $\theta + \theta' + \theta'' - 2 \pi < 2 \pi$ for both $h(gg',g'')$ and $h(g,g'g'')$, we obtain $h (gg',g'') = 0$ and $h(g,g'g'') = 0$.

\item[(g)] From $\theta + \theta' < 2 \pi$ and $\theta' + \theta'' < 2 \pi$, we obtain $h(g,g')=0$ and $h(g',g'')=0$. On the other hand, since $\theta + \theta' + \theta''  \ge 2 \pi$ for both $h(gg',g'')$ and $h(g,g'g'')$, we obtain $h (gg',g'') = 1$ and $h(g,g'g'') = 1$.

\item[(h)] From $\theta + \theta' < 2 \pi$ and $\theta' + \theta'' < 2 \pi$, we obtain $h(g,g')=0$ and $h(g',g'')=0$. On the other hand, from $\theta + \theta' + \theta'' \ge 2 \pi$, we obtain $h (gg',g'') = 1$ and $h(g,g'g'') = 1$.  

\item[(i)]  From $\theta + \theta' < 2\pi$ and $\theta' + \theta'' < 2\pi $, we obtain $h(g,g') = 0$ and $h(g',g'') =0$, respectively. We also have $h(gg',g') = h(g,g'g'') =0$ because of $\theta + \theta' + \theta'' < 2 \pi$.
\end{itemize}

Next, we show that there exists the unique inverse of $(k,g) \in \mathbb{Z} \times_h K'$.  We assume that $(l,g')$ and $(m,g'')$ are the inverse elements of $(k,g)$ such that
\begin{subequations}
 \begin{align}
  (k,g) \cdot (l,g') &= (k+l + h(g,g'), gg') = (0, e), \label{hprod-1-a} \\
  (k,g) \cdot (m,g'') &= (k+m + h(g,g''), gg'') = (0, e), \label{hprod-1-b}
 \end{align}
\end{subequations}
where $(0,e)$ is the identity element of $\mathbb{Z} \times_h K$.  From Eqs. (\ref{hprod-1-a}) and (\ref{hprod-1-b}), we obtain $g'= g^{-1}$ and $g'' = g^{-1}$. Thus, we get $g' = g''$ and $h(g,g') = h(g,g'') = h(g, g^{-1}) = 0$. Moreover, since $k+l=0$ and $k+m=0$, we obtain $l= m = -k$. Therefore, the inverse of $(k,g)$ is only $(-k,g^{-1})$.  
\section{The conjugacy classes of the fundamental group in the cyclic and BN phases}
\label{app-b}
In the spin-2 BEC, the cyclic and BN phases accommodate many different types of vortices, which can be distinguished by the conjugacy classes of the fundamental group.  Here, we enumerate the conjugacy classes of the fundamental group in both cases. First, the fundamental group in the cyclic phase is given by 
\begin{align}
\pi_1 ((U(1)_{\phi} \times S^3_{\bm{F}})/T^{\ast}_{\phi,\bm{F}},(1,x_0)) \cong \mathbb{Z} \times_h T^{\ast}_{\phi,\bm{F}}. \label{conj-1}
  \end{align}
  From Eq. (\ref{app:sp-16}), the elements of $ \mathbb{Z} \times_h T^{\ast}_{\phi,\bm{F}}$ are given by
\begin{multline}
  \{ (n,(1,\pm \bm{1})) ,  (n, (1, \pm i \sigma_x)) , (n, (1,\pm i  \sigma_y) ) , (n, (1,\pm i  \sigma_z) ), \\  (n, (e^{  \frac{2 \pi i}{3} }, \pm C_3)  ) , (n, (e^{  \frac{2 \pi i}{ 3} },\pm i \sigma_x C_3) ) , (n, (e^{  \frac{ 2 \pi i}{ 3} }, \pm i \sigma_y C_3 ) ),(n, (e^{  \frac{2 \pi i}{ 3} }, \pm i \sigma_z C_3)), \\ (n, (e^{  \frac{- 2 \pi i}{ 3} }, \pm C_3^2) ), (n, (e^{ \frac{-2 \pi i}{ 3}},\pm i \sigma_x C_3^2 )) , (n, (e^{ \frac{-2 \pi i}{ 3}}, \pm i \sigma_y C_3^2)), (n, (e^{ \frac{-2 \pi i}{ 3}}, \pm i \sigma_z C_3^2 ) \}, \label{conj-2}
\end{multline}
  where their product is definded in Eq. (\ref{S3-2-9}). The conjugacy classes of Eq. (\ref{conj-2}) are given by \cite{Semenoff;2007} 
\begin{multline}
  \{ \{ (n, (1, \bm{1}))\}, \{ ( n,(1,-\bm{1}))\}, \{  (n, (1, \pm i \sigma_x) ), (n, (1, \pm i  \sigma_y) ), (n, (1, \pm i  \sigma_z) )\}, \\ \{ (n , (e^{ \frac{2 \pi i}{3} },  C_3)  ), (n, (e^{ \frac{2 \pi i}{3} }, - i \sigma_x C_3 ) , (n, (e^{ \frac{2 \pi i}{3} }, - i \sigma_y C_3 )), (n, (e^{ \frac{2 \pi i}{3} }, - i \sigma_z C_3 ) ) \}, \\ \{ (n, (e^{ \frac{2 \pi i}{3} }, - C_3 ) ) , (n, (e^{ \frac{2 \pi i}{3} }, i \sigma_x C_3 ) ) , (n, (e^{ \frac{2 \pi i}{3} },  i \sigma_y C_3)),  (n, (e^{ \frac{2 \pi i}{3} }, i \sigma_z C_3 )  ) \}, \\ \{ (n, (e^{ \frac{-2 \pi i}{3} }, C_3^2) ), (n,  (e^{ \frac{-2 \pi i}{3} }, i \sigma_x C_3^2 ) ) ,  (n, (e^{ \frac{-2 \pi i}{3} }, i \sigma_y C_3^2 )  ) ,  (n,  (e^{ \frac{-2 \pi i}{3} }, i \sigma_z C_3^2 ) ) \} , \\  \{ (n,(e^{ \frac{-2 \pi i}{3} }, - C_3^2 ) ), (n, (e^{ \frac{-2 \pi i}{3} }, - i \sigma_x C_3^2 ) ) ,(n,  (e^{ \frac{-2 \pi i}{3} }, - i \sigma_y C_3^2 ) ) , (n, (e^{ \frac{-2 \pi i}{3} },  - i \sigma_z C_3^2) )\}\}. \label{conj-3}
\end{multline} 
Hence, we obtain seven different conjugacy classes for the fixed $n$.

Next, we consider the BN phase. The fundamental group is given by 
\begin{align}
\pi_1 ((U(1)_{\phi} \times S^3_{\bm{F}})/(D_4^{\ast})_{\phi,\bm{F}},(1,x_0)) \cong \mathbb{Z} \times_h ( D_4^{\ast})_{\phi,\bm{F}}. \label{conj-4}
  \end{align}
From Eq. (\ref{app:sp-16}), elements of $ \mathbb{Z} \times_h (D_4^{\ast})_{\phi,\bm{F}}$ are represented as  
\begin{multline}
 \{ (n, (1, \pm \bm{1}) ), (n, (1, \pm i \sigma_x))  , (n, (1, \pm i \sigma_y ) ) ,(n, (1, \pm  i \sigma_z ) ) , \\ (n, ( e^{ i \pi },\pm C_4))  , (n, ( e^{ i \pi },\pm C_4^3))   , (n,( e^{ i \pi },\pm i \sigma_x C_4)) , (n ,( e^{ i  \pi }, \pm i \sigma_x C_4^3) ) \}. \label{conj-5}
\end{multline}
where their product is definded in Eq. (\ref{S3-2-9}). The conjugacy classes of Eq. (\ref{conj-5}) is given by
\begin{align}
&\{ \{ (n,(1,\bm{1})) \},\{ (n,(1,-\bm{1})) \} , \{ (n,(1,\pm i \sigma_z )) \} , \notag \\ &\{ (n,(1,\pm i \sigma_x )) , (n,(1,\pm i \sigma_y )) \} , \{ (n,(e^{i \pi},C_4)),(n,(e^{i \pi},-C_4^3)) \} , \notag \\& \{ (n,(e^{i \pi},-C_4)),(n,(e^{i \pi},C_4^3)) \} , \{ (n,(e^{i \pi},\pm i \sigma_x C_4)),(n,(e^{i \pi},\pm i \sigma_x C_4^3)) \} \} . \label{conj-6}
\end{align}
Therefore, we obtain seven different conjugacy classes for the fixed $n$.


\begin{thebibliography}{999} 

 \bibitem{Mermin;1979}
 N.~D.~Mermin, Rev.\ Mod.\ Phys.\ {\bf 51}, 591 (1979).
  \bibitem{Trebin;1982}
  H.-R.~Trebin, Adv.\ Phys.\ {\bf 31}, 195 (1982).
  \bibitem{Michel;1980}
  L.~Michel, Rev.\ Mod.\ Phys.\ {\bf 52}, 617 (1980). 
 \bibitem{Mineev;1998}
 V.~P.~Mineev, {\it Topological stable Defects and Solitons in Ordered Media}, Classical Reviews in Physics, Vol.1, Harwood Accademic Publishers Amsterdam, (1998).
 \bibitem{Volovik;2003}
 G.~E.~Volovik, {\it The Universe in a Helium Droplet}, Oxford University Press, Oxford, (2003).
%\cite{Preskill:1979zi}
%\bibitem{Preskill;1979zi}
 %J.~Preskill,
 %``Cosmological Production Of Superheavy Magnetic Monopoles,''
 %Phys.\ Rev.\ Lett.\  {\bf 43}, 1365 (1979).
 %%CITATION = PRLTA,43,1365;%%
%\cite{Zeldovich:1978wj}
%\bibitem{Zeldovich:1978wj}
% Y.~B.~Zeldovich and M.~Y.~Khlopov,
 %``On The Concentration Of Relic Magnetic Monopoles In The Universe,''
% Phys.\ Lett.\  B {\bf 79}, 239 (1978).
 %%CITATION = PHLTA,B79,239;%% 
 \bibitem{Kibble;1976}
  T.~W.~B.~Kibble, J. Phys. A{\bf 9}, 1387 (1976).
  \bibitem{Zurek;1985}
  W.~H.~Zurek, Nature (London), {\bf 317}, 505 (1985).
  \bibitem{Volovik;1977inf}
 %Investigation of singularities in superfluid $^3$He-A
 G.~E.~Volovik and V.~P.~Mineev, Zh. Eksp. Teor. Fiz. {\bf 45} 1186 (1977) [Sov. Phys. JETP {\bf 45}, 1186 (1977)].
 \bibitem{Mermin;1978inf}
 N.~D.~Mermin, J. Math. Phys. {\bf 19 } 1457 (1978).
%  \bibitem{Nakanishi;1988}
% H.~ Nakanishi, K.~ Hayashi, and H.~ Mori, Comm.\ Math.\ Phys.\ {\bf 117}, 203 (1988)
  \bibitem[A.~S.~Schwarz (1982)]{Schwarz;1982}
 A.~S.~Schwarz, Nucl.\ Phys.\ B {\bf 208}, 141 (1982).
 \bibitem[M.~Bucher,H.-K.~Lo, J.~Preskill]{Bucher;1992}
 M.~Bucher,H.-K.~Lo, J.~Preskill, Nucl.\ Phys.\ B {\bf 386} 3 (1992).
 %\bibitem{Vilenkin;1994}
%A.~Vilenkin and E.~P.~S.~Shellard, ``Cosmic Strings and Other Topological Defects", Cambridge Monographs on Mathematical Physics (1994).
%\bibitem{Topological_Classification}
  \bibitem{Abe;1940}
 M.~Abe,  Jap.\ J.\ Math.\ {\bf 16}, 179 (1940).
 %\bibitem{footnote-1}
%The second Abe homotopy group is also useful to classify for a ring shape defect in [H. Nakanishi, K. Hayashi, and H. Mori, Comm.\ Math.\ Phys.\ {\bf 117}, 203 (1988)].
\bibitem{Eilenberg;1939}
 S.~Eilenberg, Fund.\ Math.\ {\bf 32}, 167 (1939).
  \bibitem{Michikazu;2009}
 M.~Kobayashi, Y.~Kawaguchi, M.~Nitta, and M.~Ueda, Phys.\ Rev.\ Lett.\ {\bf 103}, 115301 (2009).
 \bibitem{Fox;1948}
R.~ H.~ Fox, Ann.\ Math.\ {\bf 49}, 471 (1948). 
%\bibitem{Nakahara;2003}
 %M.~Nakahara, {\it Geometry, Topology and Physics second Edition }, Graduate Student Series in Physics, Institute of Physics Publishing, Bristol and Philadelphia (2003).
%\bibitem{Armstrong;1979}
% M.~A.~Armstrong, {\it Basic Topology}, Undergraduate Texts in Mathematics, Springer (1979).   
\bibitem{Greenberg:1981}
 M.~J.~Greenberg, J.~R.~Harper, {\it Algebraic Topology}, A First Course, Benjamin/Cummings, Reading, MA, 1981. (1981). 
 \bibitem{Eilenberg:1939th}
 The Hurewise map is the map $h :[\tilde{f}_{\tilde{x}_0}] \in \pi_n (S^n, \tilde{x}_0) \mapsto \tilde{f}^{\ast}_{\tilde{x}_0} [\sigma] \in H_n (S^n)$, where $[\sigma] \in H_n (I^n,\partial I^n)$ is a relative homology group of $(I^n,\partial I^n)$, $I^n$ is an $n$-dimensional cube, and $\partial I^n$ is a boundary of $I^n$. Eilenberg proved that Hurewise map $h$ commutes with $\tilde{\gamma}_g$ such that $h(\tilde{\gamma}_g(\tilde{f}_{\tilde{x}_0})) = g_{\ast} h(\tilde{f}_{\tilde{x}_0})$, where $g_{\ast}$ is an induced map of $g$. By the de Rham duality, $[\sigma]$ corresponds one-on-one to $[\omega] \in H^n (S^n)$. Therefore, we can use Eilenberg's theorem for the de Rham cohomology group. 
%\bibitem{G.Whitehead;1978}
%G.~W.~Whitehead, {\it Elements of Homotopy Theory}, Graduate Texts in Mathematics, Springer-Verlag (1978). 
%LQ%%%%%%%%%%%%%%%%%%%%%%%%%%%%%%%%%%%%%%%%%%%%%%%%%%%%%%%%%%%%%%%%%%%%%%%%%%
\bibitem{deGenne:1993}
P.~D.~Gennes and J.~Prost, {\it The Physics of Liquid Crystals}, Oxford University Press, Oxford, (1993).
%%Spinor BECs%%%%%%%%%%%%%%%%%%%%%%%%%%%%%%%%%%%%%%%%%%%%%%%%%%%%%%%%%%%%%%%%%
\bibitem{Ueda:arXiv}
M. Ueda and Y. Kawaguchi, arXiv:1001.2072
   \bibitem{Makela:2003}
  H.~M\"{a}kel\"{a}, Y.~Zhang, and K.-A.~Suminen, J. Phys. A:  Math. Gen. {\bf 36}, 8555 (2003).
  \bibitem{Yip:2007}
  S.~-K.~Yip, Phys. Rev. A {\bf 75}, 023625 (2007).
\bibitem{Ho;1998}
 %Spinor Bose Condensates in Optical Trap
 T.-L.~Ho, Phys. Rev. Lett. {\bf 81}, 742 (1998).
 \bibitem{Ohmi;1998}
 T.~Ohmi and K.~Machida, J. Phys. Soc. Jpn. {\bf 67} 1822 (1998).
  \bibitem{Khawaja;2001}
 U.~A.~Khawaja and H. Stoof, Nature (London) {\bf 411} 918 (2001).
   \bibitem{Fei:2001}
 F. Zhou, Phys. Rev. Lett. {\bf 87}, 080401 (2001). 
  \bibitem{Leonhardt;2000}
 %How to create an Alice String in Vector Bose-Einstein Condensate
 U. Leonhardt and G. E. Volovik,  Pis'ma Zh. Eksp. Teor. Fiz. {\bf 72}, 66 (2000)  [JETP. Lett.\ {\bf 72}, 46 (2000)].
 \bibitem{Stoof:2001}
 H.~T.~C.~Stoof, E.~Vliegen, and U.~Al~Khawaja, Phys. Rev. Lett. {\bf 87}, 120407 (2001). 
 \bibitem{Kawaguchi;2008}
 Y.~Kawaguchi, M.~Nitta, and M.~Ueda, Phys. Rev. Lett. {\bf 100}, 180403 (2008).
  \bibitem{Ueda;2000}
 %exact eigenvalue
 M.~Koashi and M.~Ueda, Phys. Rev. Lett. {\bf 63} 013601 (2000).
 \bibitem{Ueda;2002}
 %Theory of spin-2 BEC
 M.~Ueda and M.~Koashi, Phys. Rev. A {\bf 65} 063602 (2002).
 \bibitem{Yip;2000}
 %Phase diagrams of F=2
 C.~V.~Ciobanu, S.-K.~Yip, and T.-L.~Ho, Phys. Rev. A {\bf 61} 033607 (2002).
  \bibitem{Mermin:1974spin-2}
 N.~D.~Mermin, Phys. Rev. A {\bf 9}, 868 (1974).
 \bibitem{Song;2007}
 %Uniaxial and Biaxial Spin Nematic
 J.~L.~Song, G.~W.~Semenoff, and F.~Zhou, Phys. Rev. Lett. {\bf 98} 160408 (2007).
 \bibitem{Tuner;2007}
 %Nematic Order by Disorder in Spin-2 BECs
 A.~M.~Turner, R.~Barnett, E.~Demler, and A.~Vishwanath, Phys. Rev. Lett, {\bf 98} 190404 (2007). 
 \bibitem{Uchino;2010}
 %Quasi-Nanbu-Goldstone
 S.~Uchino, M.~Kobayashi, M.~Nitta, and M.~Ueda, Phys. Rev. Lett. {\bf 105} 230406 (2010)
\bibitem{Semenoff;2007}
 %1/3-vortex
 G.~W.~Semenoff and F.~Zhou, Phys.\ Rev.\ Lett.\ {\bf 98}, 100401 (2007).
% \bibitem{Shankar;1977}
 % R.~Shankar, J.\ phys.\ Lett. (Paris), {\bf 38}, 1405 (1977).  
 % \bibitem{Volovik;1977}
 % G.~E.~Volovik and V.~P. ~Mineev, Zh. Eksp. Teor. Fiz. {\bf 73}, 767 (1977) [Sov. Phys. JETP {\bf 46}, 401 (1977)].
 %\bibitem{Schwarz;1982}
% A.~S.~Schwarz, Nucl.\ Phys.\ B {\bf208}, 141 (1982).
 %\cite{Preskill:1979zi}
\bibitem{Volovic:1982} 
G.~E.~Volovik and M.~V.~Khazan, JETP. {\bf 55}, 867 (1982); G.~E.~Volovik and M.~V.~Khazan, JETP. {\bf 58}, 551 (1983).
\bibitem{Novikov:1982}
S.~P.~Novikov, Uspekhi Mat. Nauk, {\bf 37}, 3 (1982).
\bibitem{Brauner:2010}
T.~Brauner , Symmetry {\bf 2}, 609 (2010). 
\bibitem{Mermin:1978qng}
N.~D.~Mermin, V.~P.~Mineyev, and G.~E.~Volovik, J. Low. Temp. Phys. {\bf 33}, 117 (1978)
\bibitem{Lyuksyutov:1978}
I.~F.~Lyuksyutov, JETP. {\bf 48}, 178 (1978)
\bibitem{Penin:1996}
A.~A.~Penin, V.~A.~Rubakov, P.~G.~Tinyakov, and S.~V.~Troitsky, Phys. Lett. B {\bf 389}, 13 (1996).
\bibitem{Achucarro:2002}
A.~Ach\'{u}carro, A.~C.~Davis, M.~Pickles, and J.~Urrestilla, Phys Rev. D {\bf 66}, 105013 (2002).
\bibitem{Preskill;1979zi}
 J.~Preskill,
 %``Cosmological Production Of Superheavy Magnetic Monopoles,''
 Phys.\ Rev.\ Lett.\  {\bf 43}, 1365 (1979).
%\bibitem{J.Whitehead;1941}
% J.~H.~C.~Whitehead, Ann.\ of Math.\ {\bf 42 }, 400 (1941).
 \end{thebibliography}
\end{document}